\documentclass[letterpaper,twocolumn,amsmath,amssymb,aps,prl,showpacs]{revtex4}
\usepackage[latin1]{inputenc}
\usepackage{color}
\usepackage{float}
\usepackage{textcomp}
\usepackage{amsmath}
\usepackage{amssymb}
\usepackage{wasysym}
\usepackage{graphicx}
\usepackage{esint}
\usepackage[unicode=true,pdfusetitle,
 bookmarks=true,bookmarksnumbered=false,bookmarksopen=false,
 breaklinks=false,pdfborder={0 0 1},backref=false,colorlinks=false]
 {hyperref}
\hypersetup{
 colorlinks,linkcolor=blue,citecolor=blue,urlcolor=blue}

\makeatletter

\pdfpageheight\paperheight
\pdfpagewidth\paperwidth

\providecommand{\tabularnewline}{\\}

\@ifundefined{textcolor}{}
{%
 \definecolor{BLACK}{gray}{0}
 \definecolor{WHITE}{gray}{1}
 \definecolor{RED}{rgb}{1,0,0}
 \definecolor{GREEN}{rgb}{0,1,0}
 \definecolor{BLUE}{rgb}{0,0,1}
 \definecolor{CYAN}{cmyk}{1,0,0,0}
 \definecolor{MAGENTA}{cmyk}{0,1,0,0}
 \definecolor{YELLOW}{cmyk}{0,0,1,0}
}

\usepackage{wasysym}

\usepackage{breakurl}

\usepackage{leftidx}

\@ifundefined{textcolor}{}{%
 \definecolor{BLACK}{gray}{0}
 \definecolor{WHITE}{gray}{1}
 \definecolor{RED}{rgb}{1,0,0}
 \definecolor{GREEN}{rgb}{0,1,0}
 \definecolor{BLUE}{rgb}{0,0,1}
 \definecolor{CYAN}{cmyk}{1,0,0,0}
 \definecolor{MAGENTA}{cmyk}{0,1,0,0}
 \definecolor{YELLOW}{cmyk}{0,0,1,0}
}

\makeatother

\begin{document}

\title{Critical Delocalization of Chiral Zero Energy Modes in Graphene}

\author{Aires Ferreira$^{1}$}

\email{aires.ferreira@york.ac.uk}

\affiliation{$^{1}$Department of Physics, University of York, York YO10 5DD,
United Kingdom}

\author{Eduardo R. Mucciolo$^{2}$}

\affiliation{$^{2}$ Department of Physics, University of Central Florida, Orlando,
Florida 32816, USA}
\begin{abstract}
Graphene subjected to \emph{chiral}-symmetric disorder is believed
to host zero energy modes (ZEMs) resilient to localization, as suggested
by the renormalization group analysis of the underlying nonlinear
sigma model. We report accurate quantum transport calculations in
honeycomb lattices with in excess of $10^{9}$ sites and fine meV
resolutions. The Kubo dc conductivity of ZEMs induced by vacancy defects
(chiral BDI class) is found to match $4e^{2}/\pi h$ within 1\% accuracy,
over a parametrically wide window of energy level broadenings and
vacancy concentrations. Our results disclose an unprecedentedly robust
metallic regime in graphene, providing strong evidence that the early
field-theoretical picture for the BDI class is valid well beyond its
controlled weak-coupling regime. 
\end{abstract}


\maketitle
After more than half a century, Anderson localization remains a central
concept in condensed matter physics, with its many ramifications providing
new insights into the behavior of disordered electrons \cite{50YoAL}.
The discovery of the ``tenfold'' symmetry classes of disordered
metals \cite{Gade=00003D00003D000026Wegner91-93,Altland=00003D00003D000026Zirnbauer}---beyond
the standard threefold Wigner-Dyson classification scheme---has revealed
a surprisingly rich diagram of Anderson localization transitions,
including multifractality and critical delocalization in low dimensions
\cite{EversMirlin_RMP}.

The interest in critical quantum transport in two-dimensional (2D)
systems has been greatly amplified with the discovery of graphene,
a one-atom-thick crystal endowed with massless Dirac fermions \cite{RMPgraphene}.
The internal pseudospin of the Dirac fermions---stemming from the
honeycomb lattice structure with two sublattices---enables a rich
variety of quantum transport phenomena \cite{PeresRMP,MuccioloLewenkopf},
including minimum conductivity in the clean limit \cite{Katsnelson06},
and crossover from weak-localization---orthogonal class---to weak-antilocalization---symplectic
class---with increasing impurity potential range \cite{Ando02}.

Recently, disordered graphene in the \emph{chiral} symmetry class
has been the focus of much attention \cite{Markos12,Balseiro14,Evers_DoS_2014,Mirlin_DoS_2014}.
In chiral models defined on bipartite lattices, disordered wave functions
come in electron-hole pairs with energies $\pm E$ linked by a unitary
matrix diagonal in the sublattice space, i.e., $|\phi_{\pm}\rangle=\hat{\sigma}_{z}|\phi_{\mp}\rangle$.
A remarkable feature of the chiral class is the existence of \emph{critical}
states at the band center---zero-energy modes (ZEMs)---possessing
multifractal statistics and absence of weak localization corrections
at all orders in perturbation theory \cite{Gade=00003D00003D000026Wegner91-93}.
In graphene, the simplest realization of critical ZEMs is provided
by randomly distributed vacancies. A vacancy is a topological defect
obtained by cutting out all adjacent bonds to a given carbon site.
Vacancies drastically affect the spectrum near the Dirac point, leading
to the appearance of ZEMs with enhanced density of states (DOS) and
quasilocalized character \cite{Pereira_07,Pereira_08}, which can
be detected by scanning tunneling microscopy \cite{Ugeda2010}. Other
examples of chiral-symmetric disorder in graphene include random non-Abelian
gauge fields (ripples) \cite{HHernandoPRL09}, and resonant scatterers
(e.g., adsorbed hydrogen) \cite{Ferreira11}. Whether quantum criticality
induced by chiral disorder could explain the resilience of the minimum
conductivity of graphene to Anderson localization is an outstanding
question.

The focus of this Letter is on vacancy-induced ZEMs, recently implicated
in a controversy regarding the exact nature of the quantum transport
at the Dirac point \cite{Harju14,Mirlin_Vacancies_14,Mayou_2013,Roche_2013}.
Vacancy-defective graphene belongs to the chiral orthogonal ensemble
(class BDI in the Altland-Zirnbauer classification of random fermion
models \cite{Altland=00003D00003D000026Zirnbauer}). The vanishing
of the $\beta$-function of the effective nonlinear sigma model (NL$\sigma$M)
led Ostrovsky \emph{et\,al}. to conjecture a line of fixed points
with nonuniversal metallic conductivity of the order of the conductance
quantum $\sigma(0)\approx e^{2}/h$ \cite{Mirlin_2006,Mirlin_Vacancies_10,Mirlin_Vacancies_14}.
However, the validity of the NL$\sigma$M of the BDI class has been
questioned, as vacancies are infinitely strong scatterers, not amenable
to perturbative analysis \cite{Evers_DoS_2014}. On the other hand,
numerical evaluations of the conductivity using wave-packet propagation
methods show localization of all states $\sigma(E)\rightarrow0$,
including the ZEMs \cite{Harju14,Mayou_2013,Roche_2013}. The Gade
singularity in the DOS approaching $E\rightarrow0$ \cite{Evers_DoS_2014},
however, raises questions on the validity of the extraction of the
conductivity using wave-packet propagation methods.

In this Letter we report on accurate calculations of the longitudinal
dc conductivity in macroscopic large disordered graphene. By employing
an \emph{exact} representation of the Kubo formula in terms of Chebyshev
polynomials, we were able to extract the behavior of $\sigma(E)$
at the Dirac point with unprecedented resolution. Our results univocally
show that vacancy-induced ZEMs display critical delocalization, as
suggested by perturbative calculations based on the NL$\sigma$M \cite{Mirlin_2006,Mirlin_Vacancies_10,Mirlin_Vacancies_14}
and numerical studies of the two-terminal conductance in nanoribbons
with resonant scalar impurities \cite{Mirlin_Vacancies_10,Mirlin_Vacancies_14}.
We find a \emph{constant} conductivity over a wide range of vacancy
concentrations, 
\[
\sigma(0)=\sigma_{\textrm{ZEM}}\left(1.00\pm0.01\right)\,,\:\:\sigma_{\textrm{ZEM}}\equiv\frac{4e^{2}}{\pi h}\,.
\]
Strikingly, the ZEM conductivity is found to be robust with respect
to variations in the inelastic broadening parameter $\eta$ entering
in the disordered Green functions down to $\eta=2.5$~meV. This result
is very surprising as vacancies are the ultimate case of a strong
short-range disorder in graphene mixing $K$ and $K^{\prime}$ valleys
\cite{PeresRMP,MuccioloLewenkopf}.

\emph{The model}.---Chiral disordered graphene is modeled by the standard
tight-binding Hamiltonian of $\pi$~electrons defined on a honeycomb
lattice 
\begin{equation}
\hat{H}=-t\sum_{\langle i,j\rangle}\:\left(\hat{a}_{i}^{\dagger}\hat{b}_{j}+\hat{b}_{j}^{\dagger}\hat{a}_{i}\right)\,,\label{eq:TB-graphene}
\end{equation}
where $\langle i,j\rangle$ denotes nearest-neighbor pairs of carbon
atoms and $t=2.7$~eV is the corresponding hopping integral \cite{RMPgraphene}.
Periodic boundary conditions along zigzag and armchair directions
are employed. The vacancies---obtained by removing the corresponding
$p_{z}$ orbitals---are distributed randomly on both sublattices with
overall concentration $n_{i}$. In what follows, we briefly outline
the Chebyshev-polynomial Green function method (CPGF) used to accurately
evaluate spectral properties and response functions of real size systems.

\emph{The CPGF approach}.---The numerical evaluation of the lattice
resolvent operator $\hat{\mathcal{G}}(z)=(z-\hat{H})^{-1}$ requires
a nonzero broadening (\emph{resolution}) parameter $\eta=\textrm{Im}\, z\apprge\delta E$,
where $\delta E$ is the mean level spacing. We are interested in
the limit of small $\delta E$, where strong quantum interference
effects associated with ZEMs can be fully appreciated \cite{EversMirlin_RMP}.
Numerical evaluations of disordered lattice Green functions in the
presence of critical states are computationally highly demanding.
In Ref.~\cite{Evers_DoS_2014} a time-domain stochastic method has
been employed to extract the DOS with high resolution. Here, we evaluate
target functions directly in the energy domain by expressing Green
functions in terms of an \emph{exact} polynomial expansion. Our approach
turns out to be particularly advantageous in the calculation of the
conductivity (see below). First-kind Chebyshev polynomials $\{T_{n}(x)\}_{n\in\mathbb{N}_{0}}$
are employed due to their superior convergence properties \cite{Boyd,KPM}.
The use of Chebyshev polynomials as a basis set requires rescaling
the spectrum of $\hat{H}$ into the interval $[-1:1]$. To this end,
we scale both operators and energy variables, $\hat{H}\rightarrow\hat{h}=\hat{H}/W$,
$\epsilon=E/W$, and $\lambda=\eta/W$, where $W$ is the half-bandwidth.
With this notation the Green function admits the following representation
\begin{equation}
\hat{\mathcal{G}}(E+i\eta)=\frac{1}{W}\sum_{n=0}^{\infty}g_{n}(\epsilon,\lambda)\mathcal{T}_{n}(\hat{h})\,,\label{eq:GPGF}
\end{equation}
where $\{\mathcal{T}_{n}(\hat{h})\}$ are defined through the Chebyshev
recursion relations: $\mathcal{T}_{0}(\hat{h})=\hat{\mathbb{I}}$,
$\mathcal{T}_{1}(\hat{h})=\hat{h}$, and $\mathcal{T}_{n+1}(\hat{h})=2\hat{h}\cdot\mathcal{T}_{n}(\hat{h})-\mathcal{T}_{n-1}(\hat{h})$.
The coefficients $\{g_{n}(\epsilon,\lambda)\}_{n\in\mathbb{N}_{0}}$
are system independent and possess a simple closed form \cite{Supp_Mat}.
The CPGF expansion (\ref{eq:GPGF}) is the starting point of the accurate
calculations reported in this work.

\begin{figure}
\centering{}\includegraphics[width=0.95\columnwidth]{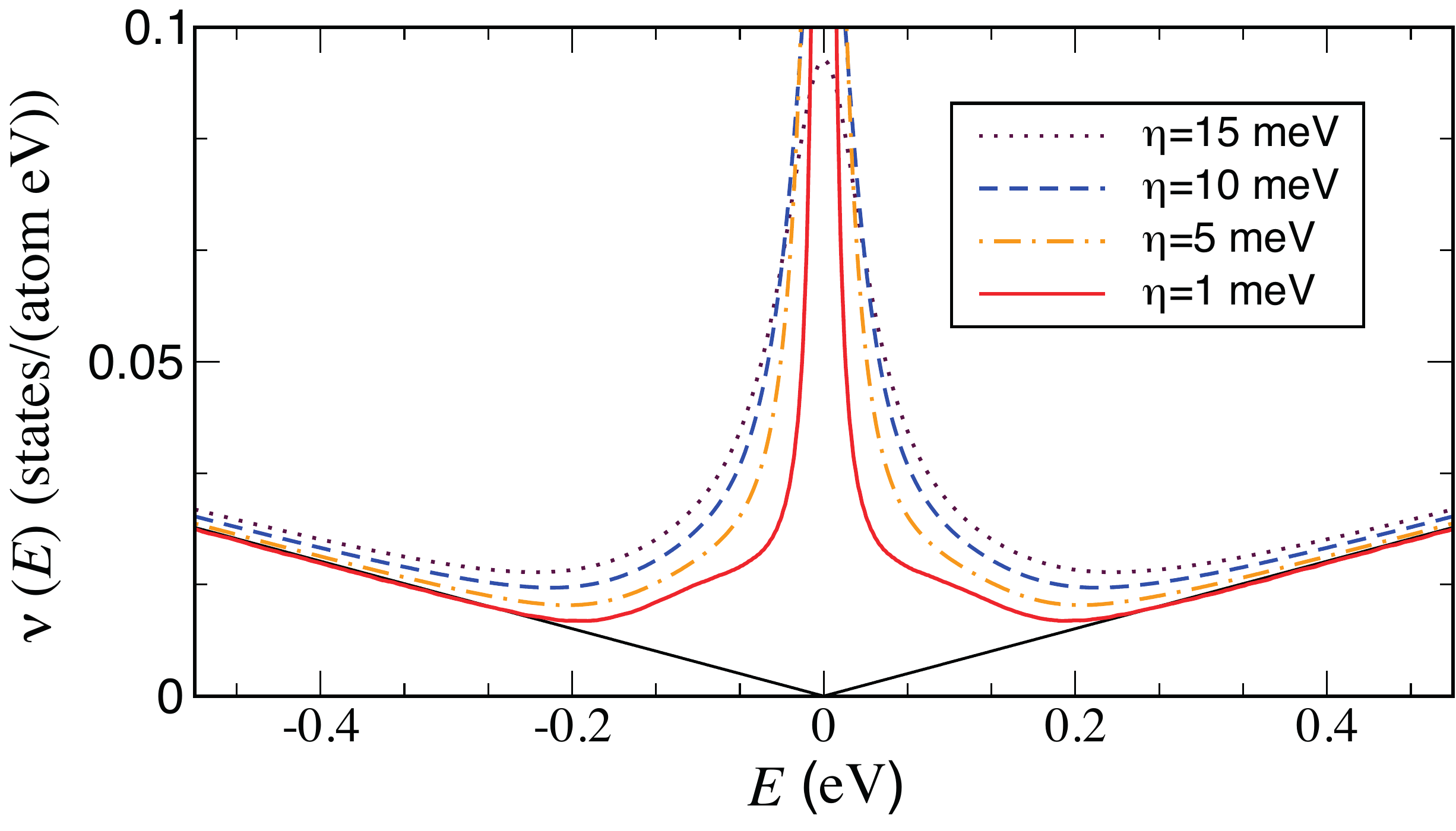} \protect\protect\protect\caption{\label{fig:01}Density of states of disordered graphene as function
of Fermi energy. The Gade singularity of ZEMs is apparent as the energy
levels are probed with increasing resolution $\eta\rightarrow0$.
The pristine DOS is shown (black line) as a guide to the eye.}
\end{figure}


\emph{Density of states}.---We start with a brief discussion of the
DOS. Formally, 
\begin{equation}
\nu(E)=-\frac{g_{s}}{\pi D}\textrm{Tr}\:\overline{\textrm{Im}\,\hat{\mathcal{G}}(E+i\eta)}\,,\label{eq:DOS}
\end{equation}
where $g_{s}=2$ accounts for spin degeneracy and the bar means disorder
averaging. According to Eqs.~(\ref{eq:GPGF})--(\ref{eq:DOS}), the
information about the DOS is contained in the Chebyshev moments $\nu_{n}=\textrm{Tr}\,\mathcal{T}_{n}(\hat{h})$
of individual disorder realizations. To probe features induced by
chiral ZEMs with meV resolution, we consider a honeycomb lattice with
$D=60\,000\times60\,000$ sites ($\approx$\,94\,$\mu$m$^{2}$).
This system has $\delta E\approx0.3$~meV at the Dirac point in the
absence of vacancies. The DOS for a dilute vacancy concentration $n_{i}=0.4\%$
is shown in Fig.~\ref{fig:01}. Given the large size of the system
simulated, one disorder configuration is sufficient to obtain very
precise results. The expected enhancement of the DOS associated with
ZEMs near $E=0$ \cite{Pereira_07,Pereira_08} is seen to dramatically
depend on the resolution. Extracting the exact scaling as $E\rightarrow0$
is a demanding task as the number of Chebyshev moments required to
converge the DOS, i.e., $N\propto W/\eta$, can be of the order of
several tens of thousands even for meV resolution; here, $N=15\times10^{3}$.
(Similar technical challenges were encountered in Ref.~\cite{Evers_DoS_2014}.)
The analysis of the data suggests that the singularity is stronger
than that predicted by Gade and Wegner \cite{Gade=00003D00003D000026Wegner91-93}
in full consistency with the detailed numerical study of Ref.~\cite{Evers_DoS_2014}
and the analytical results in Ref.~\cite{Mirlin_DoS_2014}; see Supplemental
Material for full details \cite{Supp_Mat}.

\emph{Conductivity}.---The finite-size Kubo formula reads 
\begin{equation}
\sigma(E)=\frac{2\hbar e^{2}}{\pi\Omega}\:\textrm{Tr}\left[\overline{\textrm{Im}\,\hat{\mathcal{G}}(E+i\eta)\,\hat{v}_{\parallel}\,\textrm{Im}\,\hat{\mathcal{G}}(E+i\eta)}\,\hat{v}_{\parallel}\,\right]\,,\label{eq:Kubo_Formula}
\end{equation}
where $\hat{v}_{\parallel}=[\hat{r}_{\parallel},\hat{H}]/i\hbar$
is the velocity operator (taken along the zigzag direction) and $\Omega$
is the area. Here, the broadening $\eta$ mimics the effect of uncorrelated
inelastic scattering processes, thus defining a time scale $\tau_{i}=\hbar/\eta$
for phase coherence in the system \cite{Thouless,Imry}.

\begin{figure}
\centering{}\includegraphics[width=0.95\columnwidth]{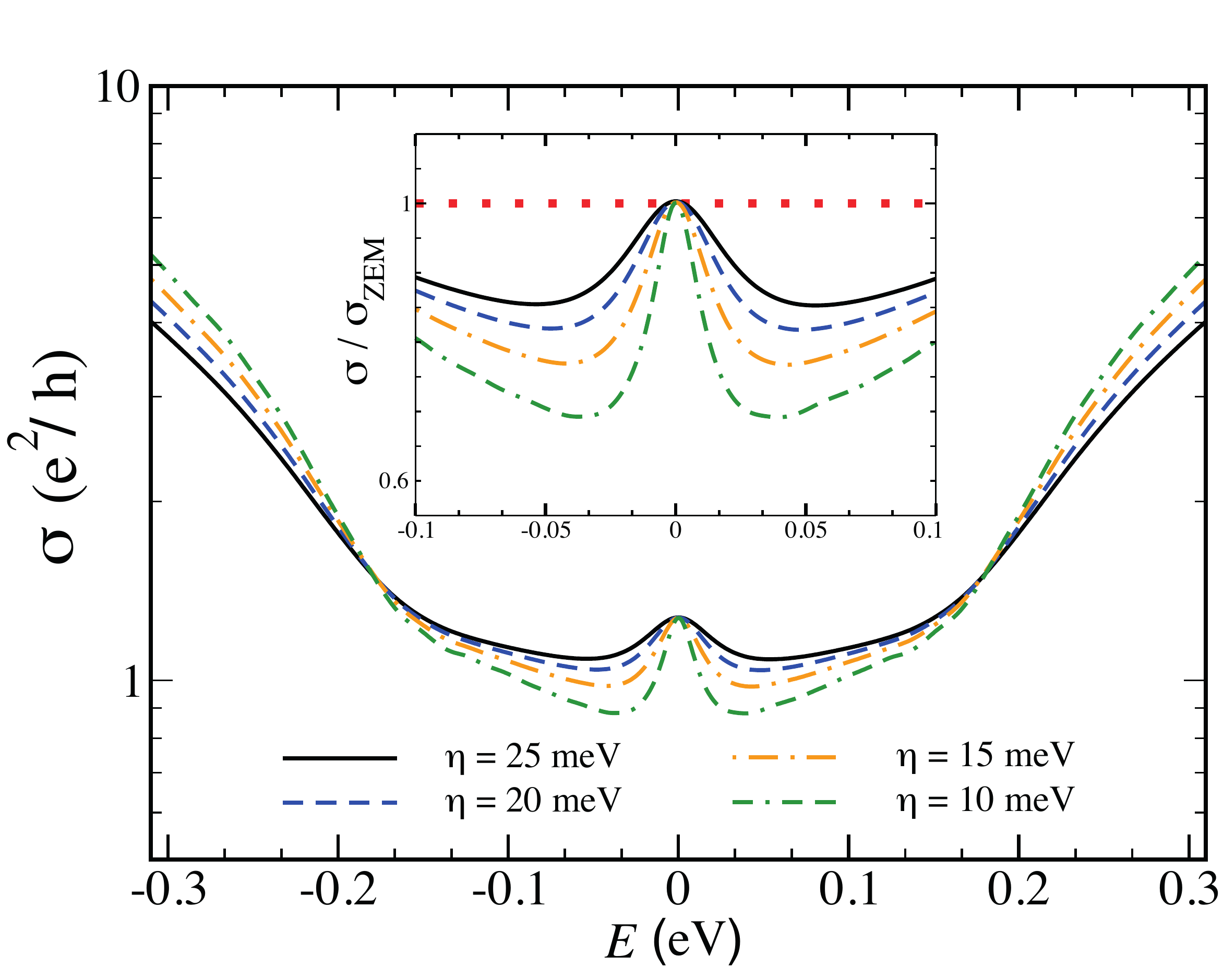} \protect\protect\protect\caption{\label{fig:02} Fully converged Kubo dc conductivity for a $0.4\%$
vacancy concentration as a function of Fermi energy at selected values
of $\eta$. The calculation required $N^{2}=6.4\times10^{7}$ Chebyshev
moments. The inset shows a zoom of the peak at the Dirac point. Statistical
fluctuations of the data are within $\simeq1$\%.}
\end{figure}

The calculation of $\sigma(E)$ follows identical steps as outlined
for the DOS. The presence of two Green functions in Eq.~(\ref{eq:Kubo_Formula})
requires a double polynomial expansion, rendering the calculation
computationally extremely demanding. Analogously to the kernel polynomial
method \cite{KPM,Ferreira11}, the full spectral information is now
contained in the Chebyshev moments $\sigma_{nm}=\textrm{Tr}\,[\mathcal{T}_{n}(\hat{h})\hat{v}_{\parallel}\mathcal{T}_{m}(\hat{h})\hat{v}_{\parallel}]$.
The number of moments required ($\equiv N^{2}$) depends on the desired
resolution. Typically, $N\approx10\times(W/\eta)$ converges the conductivity
to two decimal places. From the knowledge of $\{\sigma_{nm}\}$ the
dc conductivity $\sigma(E)$ is quickly reconstructed. See Ref.~\cite{Supp_Mat}
for details.

\emph{Full spectral} \emph{results}.---We first provide a bird's-eye
view of $\sigma(E)$ before specializing to the case of ZEMs. For
modest resolutions, $\eta\gtrsim$ 10\,meV, the physically meaningful
limit $\sigma_{\Omega\rightarrow\infty}(E)$ is achievable in relatively
small systems with $D\approx10^{7}$. The fully converged dc conductivity
for a dilute vacancy concentration $n_{i}=0.4\%$ is shown in Fig.~\ref{fig:02}.
The behavior of $\sigma_{\Omega\rightarrow\infty}(E)$ with decreasing
$\eta$ (i.e., increasing $\tau_{i}$) provides direct information
on the quantum transport regime {[}e.g., $\lim_{\eta\rightarrow0}\sigma_{\Omega\rightarrow\infty}(E)=0(>0)$
in the insulating (metallic) phase{]} \cite{Imry}. The limit $\Omega\rightarrow\infty$
is implicit hereafter. In an energy window $\simeq\pm0.2$~eV around
$E=0$---excluding the Dirac point itself---$\sigma(E)$ \emph{decreases}
as $\eta$ is lowered, showing that localization effects become increasingly
more important as the thermodynamic limit $\eta\rightarrow\delta E\rightarrow0$
is approached. The effect is notably stronger in the vicinity of the
Dirac point, where strong localization ($\sigma\lesssim e^{2}/h$)
takes place already for $\eta\approx$10~meV. This indicates that
the \emph{a priori} \emph{unknown} simulated inelastic lengths $L_{i}=L_{i}(E,\tau_{i})$
are sufficiently large that charge carriers can effectively experience
localization. In contrast, at energies $|E|\gtrsim0.2$~eV an \emph{increase}
of $\sigma(E)$ with increasing $\tau_{i}$ is observed. This suggests
that at such energies the simulated $L_{i}$ is not yet sufficiently
large to observe localization effects. This interpretation is further
confirmed below. At the Dirac point, on the other hand, $\sigma(E)$
seems insensitive to the inelastic broadening parameter, matching
$\sigma_{\textrm{ZEM}}$ with 1\% precision in the entire range (see
inset to Fig.~\ref{fig:02}). The anomalous robustness of the dc
conductivity as $E\rightarrow0$ is highly suggestive of a quantum
critical point, in agreement with field-theoretical predictions \cite{Mirlin_2006}.

\emph{High resolution} \emph{results}.---To probe the extension of
delocalization effects at the Dirac point, we devise a scheme to enable
the computation of $\sigma(E)$ with meV resolution. First, we recursively
construct the vectors 
\begin{align}
|\varphi_{\pm}(E)\rangle & =\frac{1}{W}\sum_{n=0}^{\infty}\textrm{Im}\left[g_{n}(\epsilon,\lambda)\right]\hat{\mathcal{O}}_{\textrm{\ensuremath{\pm}}}^{n}|\varphi\rangle\,,\label{eq:Psi}
\end{align}
where $|\varphi\rangle=\sum_{i=1}^{D}\chi_{i}|i\rangle$ is a real
random vector, $\hat{\mathcal{O}}_{\textrm{+}}^{n}=\mathcal{T}_{n}(\hat{h})\hat{v}_{\parallel}$,
and $\hat{\mathcal{O}}_{\textrm{-}}^{n}=\hat{v}_{\parallel}\mathcal{T}_{n}(\hat{h})$.
The random variables $\{\chi_{i}\}$ are uncorrelated and taken from
a uniform distribution with $\langle\langle\chi_{i}\rangle\rangle=0$.
The series is truncated at $n<N$ when convergence to the desired
precision is achieved. Finally, the Kubo dc conductivity is obtained
from 
\begin{equation}
\sigma_{\varphi}(E)=\frac{2\hbar e^{2}}{\pi\Omega}\:\langle\varphi_{-}(E)|\varphi_{+}(E)\rangle\,,\label{eq:Cond_SE}
\end{equation}
by averaging with respect to both disorder and random vector realizations,
i.e., $\sigma(E)=\langle\langle\overline{\sigma_{\varphi}(E)}\rangle\rangle$
\cite{Supp_Mat}. We note that for ZEMs, Eq.~(\ref{eq:Psi}) acquires
a particular simple form, $|\varphi_{\pm}(0)\rangle=W^{-1}\sum_{n}\textrm{Im}\left[g_{2n}(0,\lambda)\right]\hat{\mathcal{O}}_{\pm}^{2n}|\varphi\rangle$.
The advantage of Eqs.~(\ref{eq:Psi}) and (\ref{eq:Cond_SE}) is
that they do not require calculation of individual Chebyshev moments
$\{\sigma_{nm}\}$ (cost $\propto N^{2}$). In practice, this allows
us to reach fine resolution (higher $N$) and also much larger systems
containing up to a few billion lattice sites \cite{Resources}.

\begin{figure}
\centering{}\includegraphics[width=0.95\columnwidth]{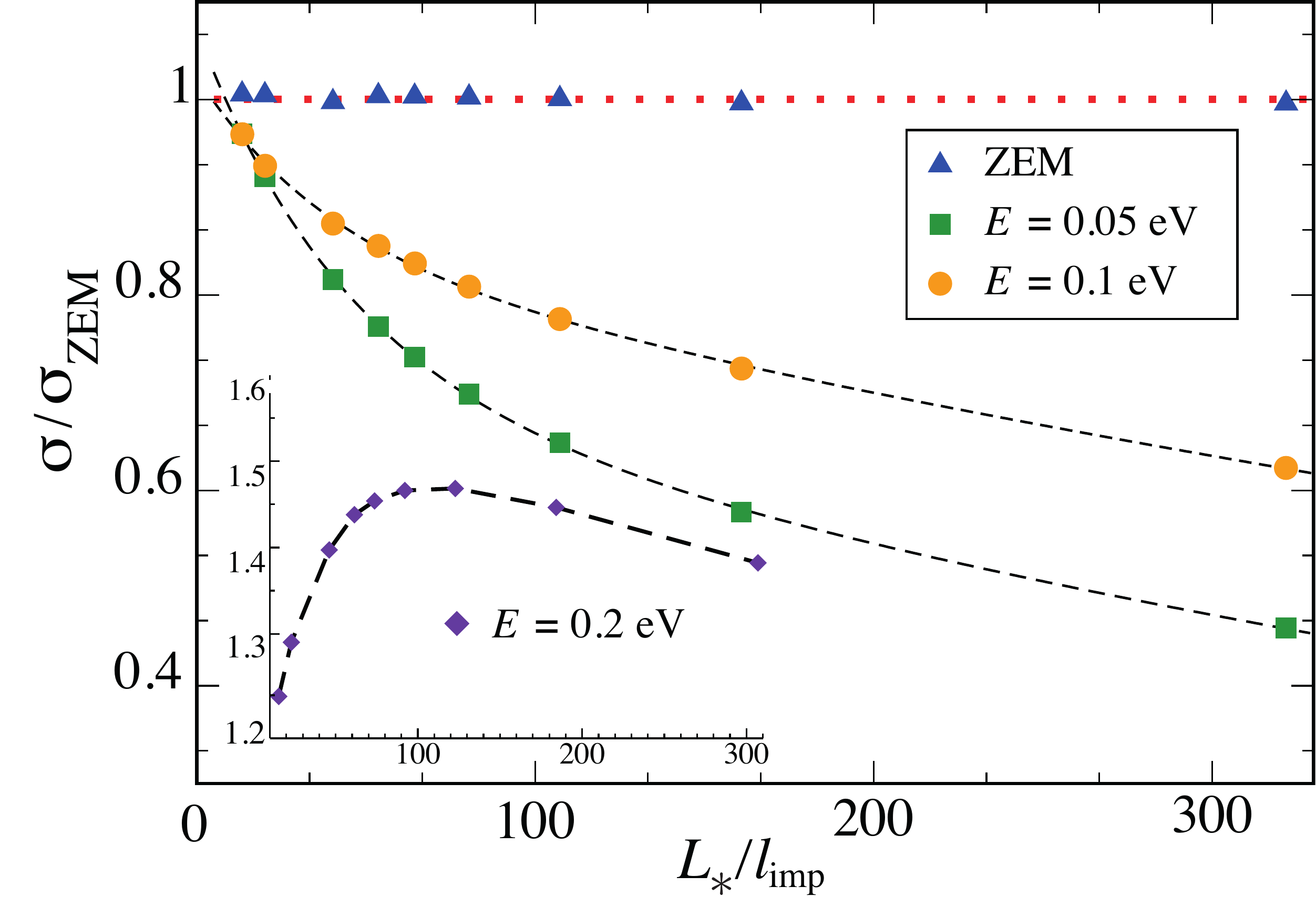} \protect\protect\protect\caption{\label{fig:03} Fully converged Kubo dc conductivity for a $0.4\%$
vacancy concentration as a function of $L_{*}/l_{\textrm{imp}}$ at
selected energies. Here $l_{\textrm{imp}}\simeq2.24$~nm is the average
distance between vacancies. A large honeycomb lattice with $3.6\times10^{9}$
sites was simulated to obtain good precision at large $L_{*}$. Statistical
fluctuations of the data are within $\simeq1$\%.}
\end{figure}


The high-resolution conductivity data across the various transport
regimes identified earlier is given in Fig.~\ref{fig:03}. For convenience,
we define an \emph{effective} system size $L_{*}\equiv\hbar\pi v_{F}/\eta$
as the length of a pristine graphene system having $\delta\epsilon=\eta$
at the Dirac point. The largest simulation has $L_{\textrm{*}}\simeq0.7$~$\mu$m,
corresponding to a broadening of only $2.5$~meV. The state vectors
in Eq.~(\ref{eq:Psi}) were calculated with $N=12\,000$ Chebyshev
iterations. The ZEM conductivity shows no sign of localization, being
numerically very close to $\sigma_{\textrm{ZEM}}=4e^{2}/(\pi h)$
through a parametrically wide range of inelastic broadenings in the
range $[2.5,60]$ meV. This is to be contrasted with the behavior
of $\sigma(E)$ away from the band center. For instance, at energies
$E=\{50,100\}$ meV there is a strong suppression towards $\sigma\rightarrow0$
as $L_{*}$ increases. The localization is stronger in the neighborhood
of the critical point at zero energy, with states with $E=50$~meV
localizing first than those having $E=100$~meV. This behavior can
also be inferred from Fig.~\ref{fig:02}, which shows that the tendency
as $\eta\rightarrow0$ ($L_{*}\rightarrow\infty$) is for states to
localize first in the vicinity of the ZEMs. In the inset to Fig.~\ref{fig:03}
the behavior for an energy far away from the Dirac point is shown.
A transition from ballistic to localized regime is observed as $L_{*}$
increases. Eventually, as $L_{*}\rightarrow\infty$, all states with
$E\neq0$ become localized. The latter is consistent with the behavior
expected for random fermions in the BDI class \cite{50YoAL,EversMirlin_RMP}.
Crucially, however, our accurate numerical treatment shows that the
chiral symmetry at $E=0$ protects ZEMs from localization up to $L_{*}\approx1$~$\mu$m.
This exotic 2D metallic regime had been predicted by the renormalization
group (RG) analysis of the NL$\sigma$M for the BDI class \cite{Mirlin_2006},
although a fully nonperturbative calculation of the microscopic conductivity
able to capture strong quantum interference effects at the Dirac point
was lacking until now.

\begin{figure}
\centering{}\includegraphics[width=0.95\columnwidth]{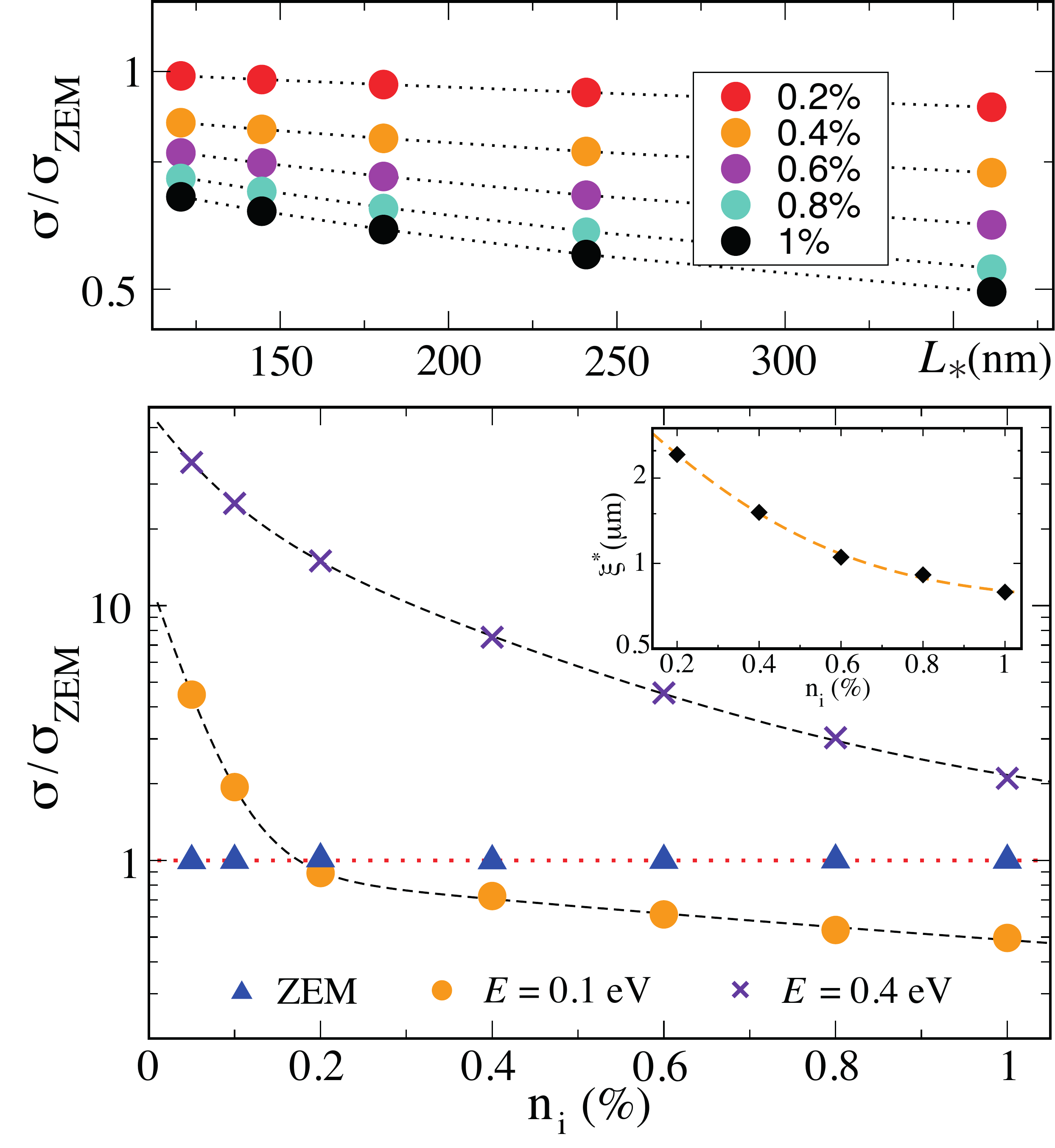} \protect\protect\protect\caption{\label{fig:04} Impact of vacancy concentration on bulk transport.
Top panel: localization of states with $E=0.1$~eV as a function
of $L_{*}$ at various vacancy concentrations. Bottom panel: variation
of $\sigma(E)$ with $n_{i}$ at selected energies.}
\end{figure}


\emph{Universal ZEM conductivity}.---We finally investigate the robustness
of the ZEMs metallic conductivity against changes in vacancy concentration.
According to the perturbative RG analysis for white-noise disorder
in the BDI class, $\sigma(0)$ should depend weakly on the disorder
strength \cite{Mirlin_2006}. The actual picture for vacancies---being
infinitely strong scatterers---is difficult to predict based solely
on field-theoretical methods \cite{Evers_DoS_2014,MirlinPRB85}. The
little sensitivity of $\sigma(0)$ to the effective length $L_{*}$
intuitively suggests a small dependence with the defect concentration
too. Interestingly, numerical results for transport across narrow
graphene strips show $\sigma(0)\approx\sigma_{\textrm{ZEM}}$ with
weak dependence on $n_{i}$ \cite{Mirlin_Vacancies_10}, demonstrating
that, although evanescent modes are strongly affected by scattering
from vacancy defects, the large number of modes available (large DOS)
counteracts perfectly to restore graphene's clean ballistic conductivity
\cite{Katsnelson06}. To investigate the possibility of a disorder-induced
universal metallic regime in graphene, we perform accurate Kubo calculations
over 2 orders of magnitude in $n_{i}$. We take a fine broadening
$\eta=2.5$ meV so as to guarantee that $L_{*}$ is sufficiently large
to capture any marked localization trend near the Dirac point. Our
results are summarized in Fig.~\ref{fig:04}. Away from the band
center the conductivity is strongly decaying with $n_{i}$ as expected.
For instance, at $E=0.1$~eV---a typical Fermi energy in experiments---the
conductivity swiftly enters in the strong localized regime already
for dilute concentrations $n_{i}\approx0.2$\%. The dependence of
$\sigma(E)$ with $L_{*}$ is well fitted by an exponential law $\sigma\propto e^{-L_{*}/\xi_{*}}$;
see top panel. (The dependence of $\xi_{*}$ with the defect concentration
is shown in the inset to the bottom panel.) However, at the band centre
ZEMs show no signs of localization even beyond the very dilute limit
up to concentrations $n=1$\%. For completeness we provide the results
for $E=0.4$~eV where transport is ballistic in the simulated range
of $L_{*}$ up to $n\approx0.8$\% (see also Fig.~\ref{fig:03}).

\textcolor{black}{We briefly comment on previous wave-packet propagation
calculations reporting on $\sigma(0)\rightarrow0$ \cite{Harju14,Mayou_2013,Roche_2013}.
The strong singularity of the DOS at $E=0$ makes the numerical extraction
of the conductivity from the Einstein relation for diffusive transport
$\sigma(E)\propto\nu(E)\, D(E)$ very challenging. Additionally, the
level broadening inserted as the inverse of the time cutoff in the
wave packet propagation may not be equivalent to the broadening employed
in the finite-size Kubo formula {[}Eq.~(\ref{eq:Kubo_Formula}){]}.
Although computationally much more demanding, our approach has the
advantage of assessing directly the microscopic conductivity with
no further assumptions.}

In summary, we have demonstrated critical delocalization of zero energy
modes in graphene by means of accurate numerical evaluations of the
Kubo conductivity in real size disordered systems containing billions
of carbon atoms. Rather remarkably, the absence of localization in
the BDI class at the Dirac point is consistent with nonlinear sigma
model predictions \cite{Mirlin_2006} and numerical studies of the
Dirac equation \cite{Mirlin_Vacancies_14,Mirlin_Vacancies_10}, suggesting
an unprecedentedly robust metallic state in two dimensions. We hope
that our work further encourages the use of accurate large-scale polynomial
methods in the study of Anderson localization transitions.

A.\,F. acknowledges M.\,D. Costa for technical discussions and high-performance
computing (HPC) support. The calculations were performed in HPC facilities
based at the Graphene Research Centre, National University of Singapore.
A.\,F. is thankful for the partial support from the National Research
Foundation, Prime Minister Office, Singapore, under its Competitive
Research Programme (Grant No. R-144-000-295-281). A.\,F. gratefully
acknowledges the financial support of the Royal Society (U.K.) through
a Royal Society University Research Fellowship. 


\newpage{}

\clearpage{}\setcounter{page}{1}

\begin{center}
\textbf{\large{}{}{}{}{}{{Supplemental Material for ``}}Critical
delocalization of chiral zero energy modes in graphene{}{}''}{\large{}{}{}{}{}{{{}
\setcounter{equation}{0}}}} {\large{}{}{}{}{}{{\setcounter{figure}{0}}}} 
\par\end{center}

\author{Aires Ferreira$^{1}$}

\affiliation{$^{1}$Department of Physics, University of York, York YO10 5DD,
United Kingdom}

\author{Eduardo R. Mucciolo$^{2}$}

\affiliation{$^{2}$ Department of Physics, University of Central Florida, Orlando,
Florida 32816, USA\bigskip{}
 }

We provide details on the Chebyshev-polynomial Green function (CPGF)
method as well as a thorough description of the accurate large-scale
numerical calculations presented in the main text.


\section{Chebyshev-Polynomial Green Function (CPGF) Method }

\label{sec:CPGF}

At the heart of the CPGF method is the exact expansion of the Green
function $\hat{\mathcal{G}}(z)=(z-\hat{h})^{-1}$ for a disordered
lattice in terms of first-kind Chebyshev polynomials \cite{CPGF-1}.
Below, we provide a short description of their main properties and
a brief derivation of the CPGF expansion. \smallskip{}

We assume that the spectrum of $\hat{h}$ falls in the interval $\mathcal{I}=[-1:1]$
\cite{rescaling}. Accordingly, in what follows, $z$ is a rescaled
complex energy variable, $z:=\epsilon+i\lambda$ with $\lambda>0$.
Chebyshev polynomials $\{T_{n}(x)\}_{n\in\mathbb{N}_{0}}$ satisfy
the recursion relations 
\begin{equation}
T_{0}(x)=1\,,\: T_{1}(x)=x\,,\: T_{n+1}(x)=2x\, T_{n}(x)-T_{n-1}(x),\label{eq:recursion relations}
\end{equation}
such that $T_{n}(x)=\cos\left(n\arccos x\right)$. They obey the orthogonality
relations 
\begin{equation}
\int_{\mathcal{I}}dx\,\omega(x)\, T_{n}(x)\, T_{m}(x)=\frac{1+\delta_{n,0}}{2}\delta_{n,m}\,,\label{eq:orthogonality_Chebyshev}
\end{equation}
where $\omega(x)=1/(\pi\sqrt{1-x^{2}})$, thus forming a complete
set in the domain $\mathcal{I}$. For a function $f(x)$ and $x\in\mathcal{I}$
one can write the expansion 
\begin{equation}
f(x)=\omega(x)\sum_{n=0}^{\infty}\frac{2\mu_{n}}{1+\delta_{n,0}}\, T_{n}(x),\label{eq:approximation_f_x}
\end{equation}
where $\mu_{n}=\int_{\mathcal{I}}dx\, f(x)\, T_{n}(x)$. Upon truncation
of the expansion, the Chebyshev polynomials distribute errors uniformly,
providing a superior polynomial expansion with uniform resolution
$\delta x\propto1/N$, where $N$ is the highest polynomial order
used \cite{Boyd-1}.

Let $\{\epsilon_{m}\}$ and $\{|m\rangle\}$ be the eigenvalues and
eigenvectors of the Hamiltonian $\hat{h}$. In order to find an exact
expansion of the lattice Green function, 
\begin{equation}
\hat{\mathcal{G}}(\epsilon+i\lambda)=\sum_{m}\frac{|m\rangle\langle m|}{\epsilon+i\lambda-\epsilon_{m}}\,,\label{eq:green_function}
\end{equation}
in terms of Chebyshev polynomials, we make use of the identity \cite{Kosloff84-1}
\begin{equation}
e^{-ixz}=\sum_{n=0}^{\infty}\frac{2i^{-n}}{1+\delta_{n,0}}J_{n}(z)\, T_{n}(x)\,,\quad|x|\le1,\label{eq:time_ev_Chebyshev}
\end{equation}
where $J_{n}(z)$ is the Bessel function of order $n$, to recast
(\ref{eq:green_function}) as 
\begin{equation}
\hat{\mathcal{G}}(\epsilon+i\lambda)=\frac{1}{i}\int_{0}^{\infty}dt\, e^{i(\epsilon+i\lambda)t}\left[\sum_{n=0}^{\infty}\frac{2i^{-n}}{1+\delta_{n,0}}J_{n}(t)\mathcal{T}_{n}(\hat{h})\right]\,,\label{eq:step_1}
\end{equation}
where $\{\mathcal{T}_{n}(\hat{h})\}$ are operators defined by the
matrix version of the Chebyshev recursion relations (\ref{eq:recursion relations}),
that is, 
\begin{equation}
\mathcal{T}_{0}(\hat{h})=\mathbb{I}_{D}\,,\;\mathcal{T}_{1}(\hat{h})=\hat{h}\,,\;\mathcal{T}_{n+1}(\hat{h})=2\hat{h}\cdot\mathcal{T}_{n}(\hat{h})-\mathcal{T}_{n-1}(\hat{h})\,,\label{eq:Chebyshev_Matrix_Recursion}
\end{equation}
with $D$ denoting the Hilbert space dimension. The Laplace transform
of the Bessel function has a well-known solution \cite{Gradshteyn-1}
\begin{equation}
\int_{0}^{\infty}dt\, e^{-st}J_{n}(t)=\frac{1}{\sqrt{1+s^{2}}}\left(\sqrt{1+s^{2}}-s\right)^{n}.\label{eq:Laplace transform}
\end{equation}
Using this expression, after the analytic continuation $s\rightarrow-iz$
and some straightforward algebra one obtains \cite{CPGF-1} 
\begin{align}
\hat{\mathcal{G}}(\epsilon+i\lambda) & =\sum_{n=0}^{\infty}g_{n}(\epsilon+i\lambda)\,\mathcal{T}_{n}(\hat{h})\,,\label{eq:CPGF}\\
g_{n}(z) & \equiv\frac{2i^{-1}}{1+\delta_{n,0}}\frac{\left(z-i\sqrt{1-z^{2}}\right)^{n}}{\sqrt{1-z^{2}}}\,.
\end{align}
In what follows we show how to use the CPGF expansion (\ref{eq:CPGF})
to compute spectral properties of large systems.


\section{Application: DOS and longitudinal dc conductivity}

\label{sec:Application_DOS_COND}

The thermodynamic density of states (DOS) is formally given by $\rho(\epsilon)=\lim_{\lambda\rightarrow0}\lim_{D\rightarrow\infty}\nu(\epsilon,\lambda)$
where 
\begin{equation}
\nu(\epsilon,\lambda)=-\frac{1}{\pi D}\:\textrm{Tr}\:\overline{\textrm{Im}\,\hat{\mathcal{G}}(\epsilon+i\lambda)}\,.\label{eq:rho_def}
\end{equation}
Here bar denotes disorder averaging. Using Eq.~(\ref{eq:CPGF}) we
easily obtain 
\begin{equation}
\nu(\epsilon,\lambda)=-\frac{1}{\pi D}\sum_{n=0}^{\infty}\textrm{Im}[g_{n}(\epsilon+i\lambda)]\,\mu_{n}\,,\label{eq:DoS_Polynom}
\end{equation}
with Chebyshev moments given by 
\begin{equation}
\mu_{n}=\textrm{Tr}[\mathcal{T}_{n}(\hat{h})]\,.\label{eq:moments_DOS}
\end{equation}
Similarly to the kernel polynomial method (KPM) \cite{KPM-1}, the
calculation of the DOS amounts to the determination of the Chebyshev
moments. This scheme is very convenient as $\{\mu_{n}\}$ can be efficiently
calculated even in very large systems with modest computational resources.
Once the moments are determined, the smeared DOS in the entire parameter
space $(\epsilon,\lambda)$ can be quickly retrieved from Eq.~(\ref{eq:DoS_Polynom}).

In a practical calculation the expansion is truncated so as to obtain
an order-$N$ approximation to the target function, 
\begin{equation}
\nu_{N}(\epsilon,\lambda)=-\frac{1}{\pi D}\sum_{n=0}^{N-1}\textrm{Im }[g_{n}(\epsilon+i\lambda)]\,\mu_{n}\,.\label{eq:smeared_DOS_N}
\end{equation}
For not too small $\lambda$, extremely accurate approximations can
be obtained for modest $N$. However, the extraction of the thermodynamic
limit almost invariably requires large $N$. The recursive calculation
of Chebyshev moments $\{\mu_{n}\}$ explore the matrix relations in
Eq.~(\ref{eq:Chebyshev_Matrix_Recursion}) and is numerically very
stable. Furthermore, the Chebyshev expansion has well defined resolution
{[}through the broadening parameter appearing in the Green function
(\ref{eq:green_function}){]}. These are two substantial advantages
of polynomial methods as compared to, e.g., Lanczos recursion \cite{KPM-1}.
Notice that kernel coefficients are absent in Eq.~(\ref{eq:smeared_DOS_N});
thus, this expansion is not equivalent to that obtained through the
KPM \cite{CPGF-1}.

The convergence rate of the exact expansion (\ref{eq:smeared_DOS_N})
depends crucially on the smoothness of the target function. As shown
by the authors in Ref.~\cite{Unified_MLG_BLG} the presence of sharp
resonances in the DOS requires a particularly large number of moments.
As a rule of thumb, the number of Chebyshev moments $N$ determine
the resolution $\delta\epsilon_{N}$ according to $\delta\epsilon_{N}\approx1/N$
\cite{KPM-1}. For instance, to probe features with small width $\eta$
an accurate calculation requires $\delta\epsilon_{N}\lesssim\eta$
and hence many Chebyshev moments before the expansion (\ref{eq:smeared_DOS_N})
converges \cite{CPGF,Unified_MLG_BLG}.

Next we discuss the application of the CPGF to the calculation of
the dc conductivity. The starting point is the finite-size Kubo formula
at zero temperature, 
\begin{equation}
\sigma(\epsilon,\lambda)=\frac{2\hbar e^{2}}{\pi\Omega}\:\textrm{Tr}\left[\,\overline{\textrm{Im}\,\hat{\mathcal{G}}(\epsilon+i\lambda)\,\hat{v}_{x}\,\textrm{Im}\,\hat{\mathcal{G}}(\epsilon+i\lambda)}\,\hat{v}_{x}\,\right],\label{eq:Kubo_formula}
\end{equation}
where $\hat{v}_{x}=[\hat{x},\hat{h}]/i\hbar$ is the velocity operator
and $\Omega$ is the area. Here, the broadening parameter $\lambda$
defines a time scale $\tau_{i}\propto1/\lambda$ for phase coherence
in the system \cite{Imry-1}. Using Eq.~(\ref{eq:CPGF}) we easily
find 
\begin{equation}
\sigma_{N}(\epsilon,\lambda)=\frac{2\hbar e^{2}}{\pi\Omega}\sum_{n,m=0}^{N-1}\textrm{Im}[g_{n}(\epsilon+i\lambda)]\textrm{Im}[g_{m}(\epsilon+i\lambda)]\,\mathcal{V}_{nm},\label{eq:Kubo_N}
\end{equation}
where 
\begin{equation}
\mathcal{V}_{nm}=\textrm{Tr}\left[\hat{v}_{x}\,\mathcal{T}_{n}(\hat{h})\,\hat{v}_{x}\,\mathcal{T}_{m}(\hat{h})\right]\,.\label{eq:moments_Kubo}
\end{equation}

The evaluation of the dc conductivity is computationally more demanding
than the DOS due to the presence of a double sum in Eq.~(\ref{eq:Kubo_N}).
The number of moments is now $N^{2}$, which can severely limit the
resolutions and/or system size attainable. However, as shown in what
follows, this limitation can be overcome if enough memory exists to
store the random vectors used for a stochastic evaluation of the moments.


\section{Efficient Calculation of Chebyshev Moments}

\label{sec:Efficient Recursive Method}

The complexity of the trace evaluation in Eqs.~(\ref{eq:moments_DOS})
and (\ref{eq:moments_Kubo}) is $\mathcal{O}(D^{2})$. However, for
very large sparse matrices, such as those appearing in effective tight-binding
models, the full trace can be replaced by a stochastic average. For
instance, for the DOS one can replace Eq.~(\ref{eq:moments_DOS})
by 
\begin{equation}
\mu_{n}\approx\frac{1}{R}\sum_{r=1}^{R}\langle r|\mathcal{T}_{n}(\hat{h})|r\rangle\,,\label{eq:STE_DOS}
\end{equation}
where $|r\rangle=\sum_{i=1}^{D}\xi_{i}|i\rangle$ are complex random
vectors with coefficients satisfying $\langle\langle\xi_{i}\rangle\rangle=0$
and $\langle\langle\xi_{i}^{*}\xi_{j}\rangle\rangle=\delta_{ij}$
(real vectors may be used for spin rotational and time reversal symmetric
Hamiltonians) \cite{Ebisuzaki_2004-1}. The number of operations required
to compute (\ref{eq:STE_DOS}) is now $\mathcal{O}(D\times R)$.

It is often assumed that the error in (\ref{eq:STE_DOS}) has the
very favorable scaling $\mathcal{O}(1/\sqrt{RD})$ \cite{KPM-1}.
However, for very large $n$ the matrix $\mathcal{T}_{n}(\hat{h})$
is no longer sparse and a larger $R$ (or a larger system size $D$)
is needed to obtain a stochastic trace evaluation (STE) with good
precision. In practice, for very large systems, with $D\approx10^{9}$,
we found that a single random vector is enough to obtain errors below
1\% for $n$ up to ten thousand. Details are given below.

We now overview the recursive method that allows us to efficiently
calculate Chebyshev moments. For a general introduction the reader
is referred to the review by A. Weisse \emph{et al}.~\cite{KPM-1}.
For concreteness, we describe the calculation of conductivity moments,
i.e., 
\begin{equation}
\mathcal{V}_{nm}=\frac{1}{R}\sum_{r=1}^{R}\langle r|\hat{v}_{x}\,\mathcal{T}_{n}(\hat{h})\,\hat{v}_{x}\,\mathcal{T}_{m}(\hat{h})|r\rangle\,.\label{eq:STE_COND}
\end{equation}
(The DOS moments are computed with a similar scheme.) Suppose we start
with a random vector $|r\rangle$. Then, using the recursion relations
{[}Eq.~(\ref{eq:Chebyshev_Matrix_Recursion}){]}, we obtain 
\begin{equation}
\mathcal{T}_{m+1}(\hat{h})|r\rangle=2\hat{h}\,\mathcal{T}_{m}(\hat{h})|r\rangle-\mathcal{T}_{m-1}(\hat{h})|r\rangle\,,\label{eq:rec_1}
\end{equation}
which inspires us to write 
\begin{equation}
|r\rangle_{m+1}=2\hat{h}|r\rangle_{m}-|r\rangle_{m-1}\,,\label{eq:stepr_2}
\end{equation}
where 
\begin{equation}
|r\rangle_{m}=\mathcal{T}_{m}(\hat{h})|r\rangle\,.\label{eq:stepr_3}
\end{equation}
In the above, $|r\rangle_{0}=|r\rangle$. In fact, the best way to
proceed is to define a second, auxiliary truncated basis $\{|\bar{r}\rangle\}$,
$\bar{r}=1,..,R$, with 
\begin{equation}
|\bar{r}\rangle=\hat{v}_{x}|r\rangle\,.\label{eq:stepr_4}
\end{equation}
Then, we can apply the Chebyshev recursion to write 
\begin{equation}
|\bar{r}\rangle_{n+1}=2\hat{h}\,|\bar{r}\rangle_{n}-|\bar{r}\rangle_{n-1}\,.\label{eq:stepr_5}
\end{equation}
The idea now is to implement a recursive calculation for each pair
of random vectors, $\{|r\rangle\}$ and $\{|\bar{r}\rangle\}$, to
generate two sequences $\{|r\rangle_{0},...,|r\rangle_{N-1}\}$ and
$\{|\bar{r}\rangle_{0},...,|\bar{r}\rangle_{N-1}\}$, since they can
be used to directly compute the stochastic averages 
\begin{equation}
\mathcal{V}_{nm}(r)=\leftidx{_{n}}{\langle\bar{r}|\hat{v}_{x}|r\rangle}{_{m}}\label{eq:Gamma_nm_r}
\end{equation}
needed for the calculation of the conductivity moments (\ref{eq:STE_COND}),
i.e., 
\begin{equation}
\mathcal{V}_{nm}=\frac{1}{R}\sum_{r=1}^{R}\mathcal{V}_{nm}(r)\,.\label{eq:COND_STE_r}
\end{equation}
If large amounts of RAM are available, one can recursively compute
$\{{|r\rangle_{n}}\}$ and $\{{|\bar{r}\text{\textrangle}_{m}}\}$
for all $n,m=0,...,N-1$, store them, and then evaluate the coefficients
$\mathcal{V}_{nm}(r)$ for each $r$ {[}no need to store $\{{|r\text{\textrangle}_{n}}\}$
and $\{{|\bar{r}\text{\textrangle}_{m}}\}$ for more than a given
$r$ at any time{]} \cite{RAM}.

We now show how to evaluate efficiently the matrix elements $\mathcal{V}_{nm}(r)=\leftidx{_{n}}{\langle\bar{r}|\hat{v}_{x}|r\rangle}{_{m}}$
using a site representation for the random vector. Let 
\begin{equation}
\psi_{n}^{(r)}(x_{k},y_{k})=\langle x_{k},y_{k}|r\rangle_{n},\quad\phi_{m}^{(r)}(x_{k},y_{k})=\langle x_{k},y_{k}|\bar{r}\rangle_{m}\,,\label{eq:s1}
\end{equation}
with $k=1,...,D$, where $(x_{k},y_{k})$ are lattice site coordinates.
Then, 
\begin{align}
\mathcal{V}_{nm}(r) & =\sum_{k,k^{\prime}=1}^{D}[\phi_{m}^{(r)}(x_{k},y_{k})]^{*}\psi_{n}^{(r)}(x_{k^{\prime}},y_{k^{\prime}})\times\nonumber \\
 & \qquad\times\langle x_{k},y_{k}|\hat{v}_{x}|x_{k^{\prime}},y_{k^{\prime}}\rangle\,.\label{eq:s2}
\end{align}
We can write 
\begin{align}
\langle x_{k},y_{k}|\hat{v}_{x}|x_{k^{\prime}},y_{k^{\prime}}\rangle & =i(x_{k^{\prime}}-x_{k})\langle x_{k},y_{k}|\hat{h}|x_{k^{\prime}},y_{k^{\prime}}\rangle\,.\label{eq:s2_aux}
\end{align}
When only nearest-neighbor hopping is allowed, there is a substantial
reduction in the number of terms required to compute the matrix element:
\begin{align}
\mathcal{V}_{nm}(r) & =i\sum_{k=1}^{D}[\phi_{m}^{(r)}(x_{k},y_{k})]^{*}\sum_{\boldsymbol{\tau}}\tau_{x}\,\psi_{n}(x_{k}+\tau_{x},yk+\tau_{y})\times\nonumber \\
 & \times\langle x_{k},y_{k}|\hat{h}|x_{k}+\tau_{x},y_{k}+\tau_{y}\rangle\,,\label{eq:s3}
\end{align}
where the number of lattice vectors $\boldsymbol{\tau}$ depends on
the topology of the problem. The number of computational steps is
thus precisely $D\times z$, which is much lower than $D^{2}$. In
most cases of interest, the Hamiltonian matrix element is just a constant
hopping amplitude $-t_{s}$, in which case we have 
\begin{equation}
\mathcal{V}_{nm}(r)=-i\, t_{s}\sum_{k=1}^{D}[\phi_{m}^{(r)}(x_{k},y_{k})]^{*}\sum_{\boldsymbol{\tau}}\tau_{x}\,\psi_{n}(x_{k}+\tau_{x},y_{k}+\tau_{y})\,.\label{eq:s4}
\end{equation}
Notice that $t_{s}$ is a dimensionless hopping amplitude since, by
construction, $||\pm\hat{h}||\le1$ (for graphene with vacancy defects,
$t_{s}\equiv t/W$ where $t$ is the carbon-carbon hopping integral
and $W$ is half-bandwidth). Clearly, there is no need to store the
entire $D\times D$ Hamiltonian matrix; a connectivity table with
information about neighbor coordinates $\{n_{1}(x_{k},y_{k}),...,n_{z}(x_{k},y_{k})\}_{k}$
suffices. This shows that the current scheme is just limited by the
memory required to store the amplitudes $\{\{\psi_{n}^{(r)}\}_{n}^{r},\{\phi_{m}^{(r)}\}_{m}^{r}\}$
needed to compute the overlap $\mathcal{V}_{nm}(r)$ for any two vectors
$\{{|r\text{\textrangle}_{n}},{|\bar{r}\text{\textrangle}_{m}}\}$.

The calculation can be made substantially more efficient if we are
just interested in evaluating the conductivity in a small rectangular
parametric grid $\{\{\epsilon_{p}\}\times\{\lambda_{q}\}\}$, $1\le p,q\le p_{\textrm{max},}q_{\textrm{max}}$.
The Chebyshev moments $\mathcal{V}_{nm}$ contain more information
than any such grid since they allow one to retrieve the complete spectral
conductivity according to Eq.~(\ref{eq:Kubo_N}). Recall that $\lambda$
is only limited by the number of Chebyshev iterations, $\textrm{min}\:\lambda\propto N^{-1}$,
and hence can be made arbitrary small by increasing $N$. The conductivity
for each point in the grid can be calculated efficiently using the
\emph{single-energy} \emph{algorithm} outlined in the main text. The
idea is to write the conductivity $\sigma_{N}(\epsilon_{p},\lambda_{q})$
for each pair $\{\epsilon_{p},\lambda_{q}\}$ {[}see Eq.~(\ref{eq:Kubo_N}){]}
as 
\begin{equation}
\sigma_{N}(\epsilon_{p},\lambda_{q})=\frac{2\hbar e^{2}}{\pi\Omega R}\sum_{r=1}^{R}\langle\varphi_{-}^{(r)}(\epsilon_{p},\lambda_{q})|\varphi_{+}^{(r)}(\epsilon_{p},\lambda_{q})\rangle\,,\label{eq:sigma_N_pq}
\end{equation}
where 
\begin{equation}
|\varphi_{+}^{(r)}(\epsilon_{p},\lambda_{q})\rangle=\sum_{n=0}^{N-1}\textrm{Im}[g_{n}(\epsilon_{p}+i\lambda_{q})]\hat{v}_{x}|r_{n}\rangle\label{eq:phi_p_pq}
\end{equation}
and 
\begin{equation}
|\varphi_{-}^{(r)}(\epsilon_{p},\lambda_{q})\rangle=\sum_{n=0}^{N-1}\textrm{Im}[g_{n}(\epsilon_{p}+i\lambda_{q})]|\bar{r}_{n}\rangle\,.\label{eq:phi_m_pq}
\end{equation}
Equations~(\ref{eq:phi_p_pq})-(\ref{eq:phi_m_pq}) can now be computed
iteratively with only a few vectors stored in memory (instead of $2\times N$
vectors). The substantial reduction in memory allocation has allowed
us to treat very large tight-binding systems, in excess of a billion
atoms ($D=3.6\times10^{9}$), with high resolution; see next section.


\section{Particular Case: Graphene with Random Vacancies}

\label{sec:Particular-Case:-Graphene}

In this section we provide the full numerical details of the calculations
presented in the main text.


\subsection{The DOS}

The DOS of a macroscopic large honeycomb lattice ($N_{x}=N_{y}=60000$;
periodic boundary conditions) with dilute randomly distributed vacancies
(concentration $n_{i}=0.4$\%) has been calculated using the CPGF
method and numerical implementations as described above. The $N$-order
approximation to the DOS is given by 
\begin{equation}
\nu_{N}(E,\eta)=-\frac{1}{R}\sum_{n=0}^{N-1}\sum_{r=1}^{R}\frac{\textrm{Im}[g_{n}(\epsilon+i\lambda)]}{\pi DW}\:\langle\overline{r_{n}|r_{n}}\rangle\,,\label{eq:DOS_N}
\end{equation}
where $W=3t$ is graphene's half-bandwidth and $|r_{n}\rangle$ as
defined in Eq.~(\ref{eq:stepr_3}), $E=\epsilon W$, and $\eta=\lambda W$.
The initial random vector used in the Chebyshev recursion reads as
$|r_{0}\rangle=\sum_{i=1}^{D}x_{i}|i\rangle$, where $\{x_{i}\}$
are generated from a uniform distribution on the interval $[-\sqrt{3},\sqrt{3}]$.
In such a large Hilbert space ($D=N_{x}\times N_{y}=3.6\times10^{9}$),
self averaging guarantees that a single random vector $R=1$ and one
disorder realization suffice to obtain accurate results even for fine
resolutions, that is, large $N$.

The accuracy of the stochastic evaluation of $\nu_{N}(\epsilon,\lambda)$
is illustrated with a few examples in Table~\ref{tab:DOS}. The superior
precision {[}better than 0.1\% for zero-energy modes (ZEMs) investigated
in the main text{]} is a consequence of the size of the system simulated.
We note that at larger values of the resolution parameter $\lambda$
($\eta$) the data precision improves because convergence is achieved
at smaller values of $N$ (see below).

\begin{widetext}

\begin{table}[H]
\centering{}%
\begin{tabular}{|c|c|c|c|c|}
\hline 
 & ZEM  & $E=0.05$ eV  & $E=0.10$ eV  & $E=0.20$ eV\tabularnewline
\hline 
\hline 
$S_{1}$  & \quad{}1.0782\quad{}  & \quad{}2.1507$\times10^{-2}$\quad{}  & \quad{}1.6777$\times10^{-1}$\quad{}  & \quad{}1.11326$\times10^{-2}$\quad{}\tabularnewline
\hline 
$S_{2}$  & \quad{}1.0786\quad{}  & \quad{}2.1495$\times10^{-2}$\quad{}  & \quad{}1.6764$\times10^{-1}$\quad{}  & \quad{}1.11259$\times10^{-2}$\quad{}\tabularnewline
\hline 
$S_{3}$  & \quad{}1.0784\quad{}  & \quad{}2.1501$\times10^{-2}$\quad{}  & \quad{}1.6705$\times10^{-1}$\quad{}  & \quad{}1.11310$\times10^{-2}$\quad{}\tabularnewline
\hline 
max$\:_{S_{i}}$$|\nu_{S_{i}}-\bar{\nu}|/\bar{\nu}$  & $\approx$ 0.02\%  & $\approx$ 0.03\%  & $\approx$ 0.30\%  & $\approx$ 0.35\%\tabularnewline
\hline 
\end{tabular}\protect\protect\protect\caption{\label{tab:DOS} Estimation of the data precision. DOS {[}\#states/(atom$\cdot$eV){]}
for three independent system realizations (disorder and initial random
vector $|r\rangle$), labeled $S_{1}$, $S_{2}$, and $S_{3}$, at
several energies for a resolution $\eta\equiv\lambda W$ of 1 meV.
The relative maximum deviation from the average $\bar{\nu}$ is shown
in the last row. The estimated accuracy is confirmed below using a
different approach.}
\end{table}

\end{widetext}

\begin{figure}
\begin{centering}
\includegraphics[width=1\columnwidth]{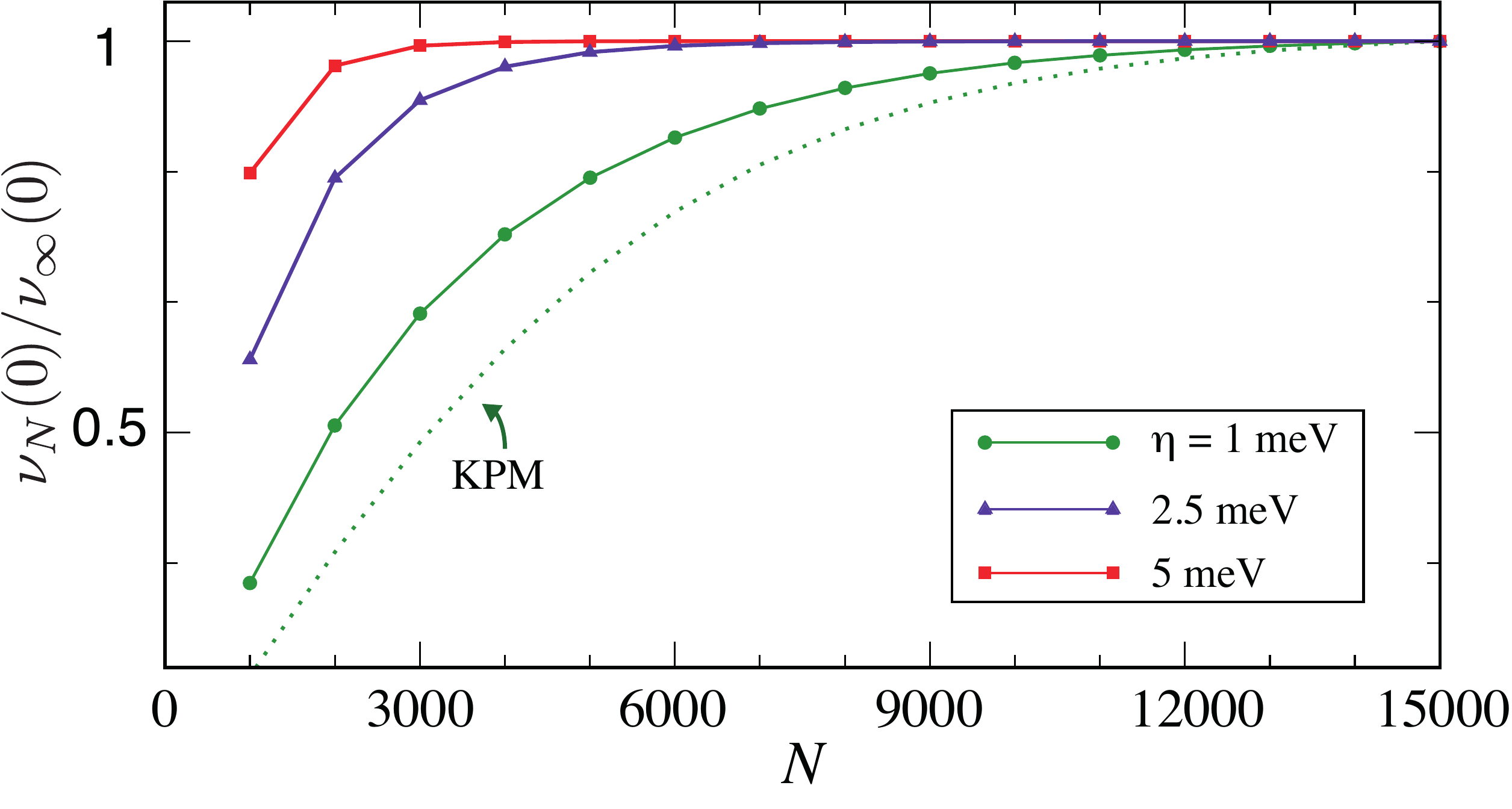} 
\par\end{centering}

\protect\protect\protect\caption{\label{fig:DOS} Convergence of the $N-$order approximation to the
DOS of ZEMs at selected values of resolution (broadening) parameter
$\eta\equiv\lambda W$ with $W=8.1$~eV. A single realization of
a disordered system with $N_{x}=N_{y}=60000$ and 0.4\% vacancy concentration
has been considered. The limiting value $\nu_{N\rightarrow\infty}(\epsilon,\lambda)$
has been estimated---with precision better than 1\%---from the value
of $\nu_{N}(\epsilon,\lambda)$ at $N=15000$. The KPM result is shown
(dotted line) for comparison.}
\end{figure}

We now assess the convergence of the $N$-order approximation. $N$
must be sufficiently large such that $\nu_{N}(\epsilon,\lambda)$
is well converged (say to 1\% accuracy or better) for the smallest
desired resolution $\lambda$. In Fig.~\ref{fig:DOS}, we show the
variation of the DOS of ZEMs $\nu_{N}(0,\lambda)$ with $N$ {[}see
Eq.~(\ref{eq:DOS_N}){]}. The calculations highlight the need for
many thousands of Chebyshev iterations when the spectrum is probed
with fine resolutions (i.e., a few meV). Similar conclusions hold
for other energies (not shown). For comparison we show the KPM approximation
to the DOS using a Lorentz kernel \cite{KPM,Unified_MLG_BLG}. Despite
being accurate in the limit $N\rightarrow\infty$, the KPM convergence
rate is manifestly poorer in this case.

\begin{figure}[H]
\begin{centering}
\includegraphics[width=1\columnwidth]{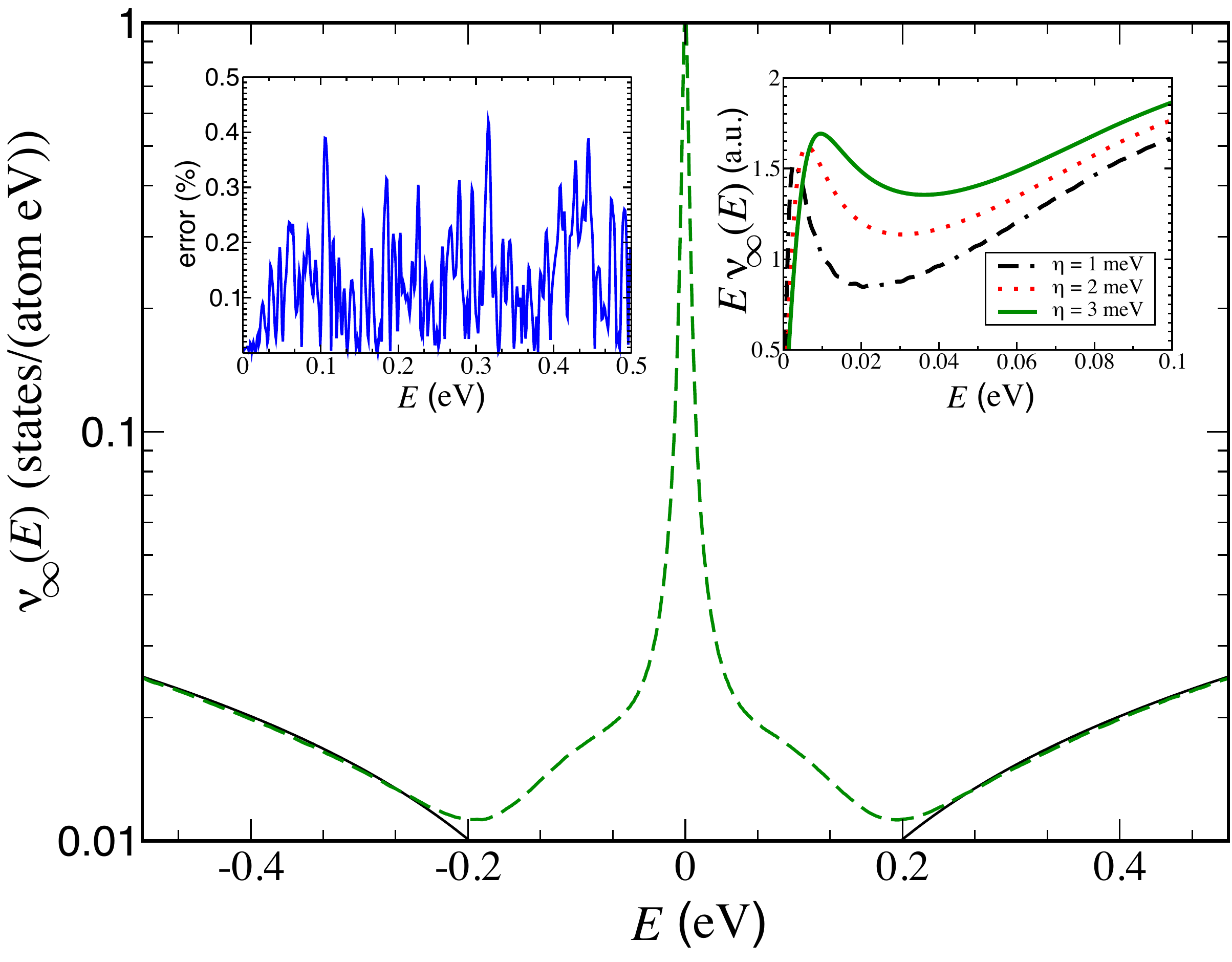} 
\par\end{centering}

\protect\protect\protect\caption{\label{fig:DOS-2} DOS of graphene with vacancy defects ($n_{i}=0.4$\%)
as function of Fermi energy (green dashed line). The resolution of
the calculation is $\eta=$1~meV. A logarithmic scale has been chosen
to highlight the singular behavior of $\nu(\epsilon,\lambda)$ as
$\epsilon\rightarrow0$. The solid black line shows the DOS of pristine
graphene as a guide to the eye. The insets show the estimated error
as function of Fermi energy (left) and a close look at the DOS singularity
at $E=0$ (right). The energy grid contains 1000 points.}
\end{figure}

Having established the convergence and accuracy of the CPGF method
in the case of graphene with vacancies, we show the fully converged
DOS for 1 meV resolution in Fig.~\ref{fig:DOS-2}. A single system
realization and random vector was employed. As an independent error
estimator we use the electron-hole asymmetry degree, i.e., $|\nu_{\infty}(\epsilon,\lambda)-\nu_{\infty}(-\epsilon,\lambda)|/\nu_{\infty}(\epsilon,\lambda)$.
The magnitude of the error and its dependencies with the Fermi energy
are consistent with the earlier statistical analysis.

In order to illustrate the divergent behavior of the DOS at $E=0$
we show in the right inset (Fig.~\ref{fig:DOS-2}) a plot of $E\nu(E,\eta)$
at several values of the resolution. According to the standard nonlinear
sigma model picture \cite{GadeWegner91-93}, the thermodynamic DOS
behaves as $\nu(E,0)\rightarrow|E|^{-1}\exp[-|\ln|E||^{-1/2}]$ as
$|E|\rightarrow0$, whereas Häfner and co-workers observed a stronger
singularity $\nu(E,0)\rightarrow|E|^{-1}|\ln|E||^{-x}$ with $2>x\ge1$
\cite{Evers_2014} in consistency with a recent prediction \cite{Mirlin_2014}.
Our results indicate $E\nu(E,\eta)\rightarrow0$ for $\eta$ down
to 1~meV, which is consistent with the numerical analysis of Ref.~\cite{Evers_2014}.
A more detailed analysis would be needed to reveal the exact dependence
as obtained in the CPGF.

Our results show that the accurate determination of the spectral properties
of disordered graphene is highly demanding, especially near the Gade
singularity where fine resolutions are needed to capture the correct
behavior. Similar challenges were reported in Ref.~\cite{Evers_2014}
where a time-domain stochastic method was used to extract the DOS.
Finally, we note that the calculations are not sensitive to the system
dimension as long as the mean level spacing is the smallest energy
scale $\delta\epsilon\lesssim\lambda$. This makes the CPGF a convenient
tool to extract the thermodynamic limit. In what follows, we show
how the CPGF behaves for the calculation of the Kubo formula.


\subsection{dc conductivity}

Below we provide the numerical details of the transport calculations
presented in the main text. In Sec.~B1 we focus on the full-spectrum
algorithm used to produce the $\sigma$ versus $E$ curve in Fig.~2
(main text). Details of the high-resolution calculations with $D=3.6\times10^{9}$
and $N$ up to 12000 {[}Figs.~3 and 4 (main text){]} are given in
Sec.~B2.


\subsubsection{Full spectral results}

As discussed in Sec.~\ref{sec:Application_DOS_COND}, the knowledge
of individual Chebyshev moments $\mathcal{V}_{nm}=\textrm{Tr}\:[\hat{v}_{x}\mathcal{T}_{n}(\hat{h})\hat{v}_{x}\mathcal{T}_{m}(\hat{h})]$
enables the full spectral determination of the dc conductivity. However,
an efficient numerical implementation requires enough memory to store
$2\times N$ vectors of dimension $D$ {[}see Eqs.~(\ref{eq:Gamma_nm_r})-(\ref{eq:Gamma_nm_r}){]},
which in practice limits the attainable $D$ and/or $N$. To boost
the size of the simulations we implemented the Chebyshev recursive
method (Sec.~\ref{sec:Efficient Recursive Method}) in machines with
large RAM. Having the sequences $\{|r\rangle_{0},...,|r\rangle_{N-1}\}$
and $\{|\bar{r}\rangle_{0},...,|\bar{r}\rangle_{N-1}\}$ stored in
RAM allows for a quick evaluation of the Chebyshev moments through
optimized linear algebra subroutines.

\begin{figure}[H]
\begin{centering}
\includegraphics[width=0.89\columnwidth]{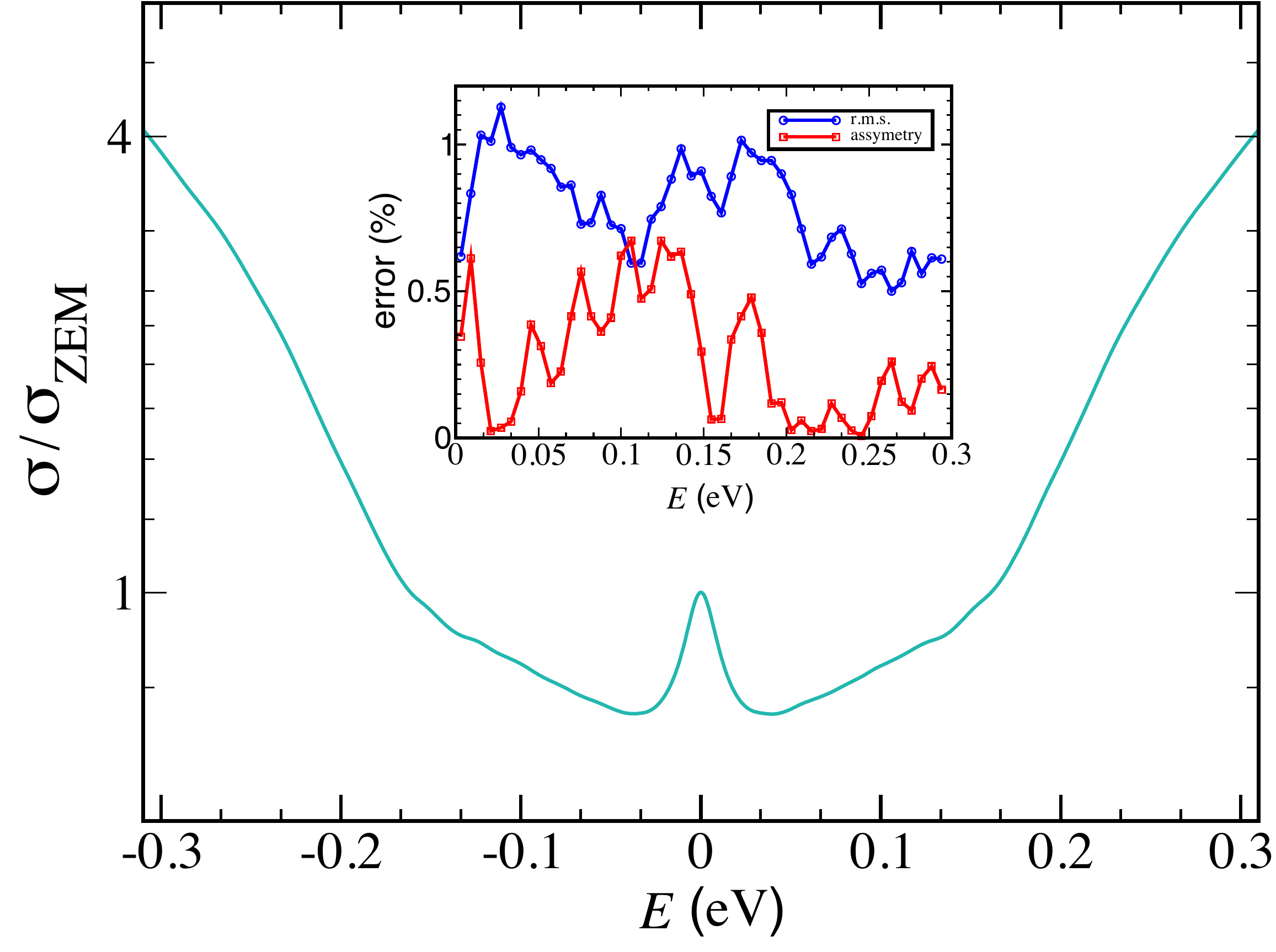} 
\par\end{centering}

\begin{centering}
\includegraphics[width=0.93\columnwidth]{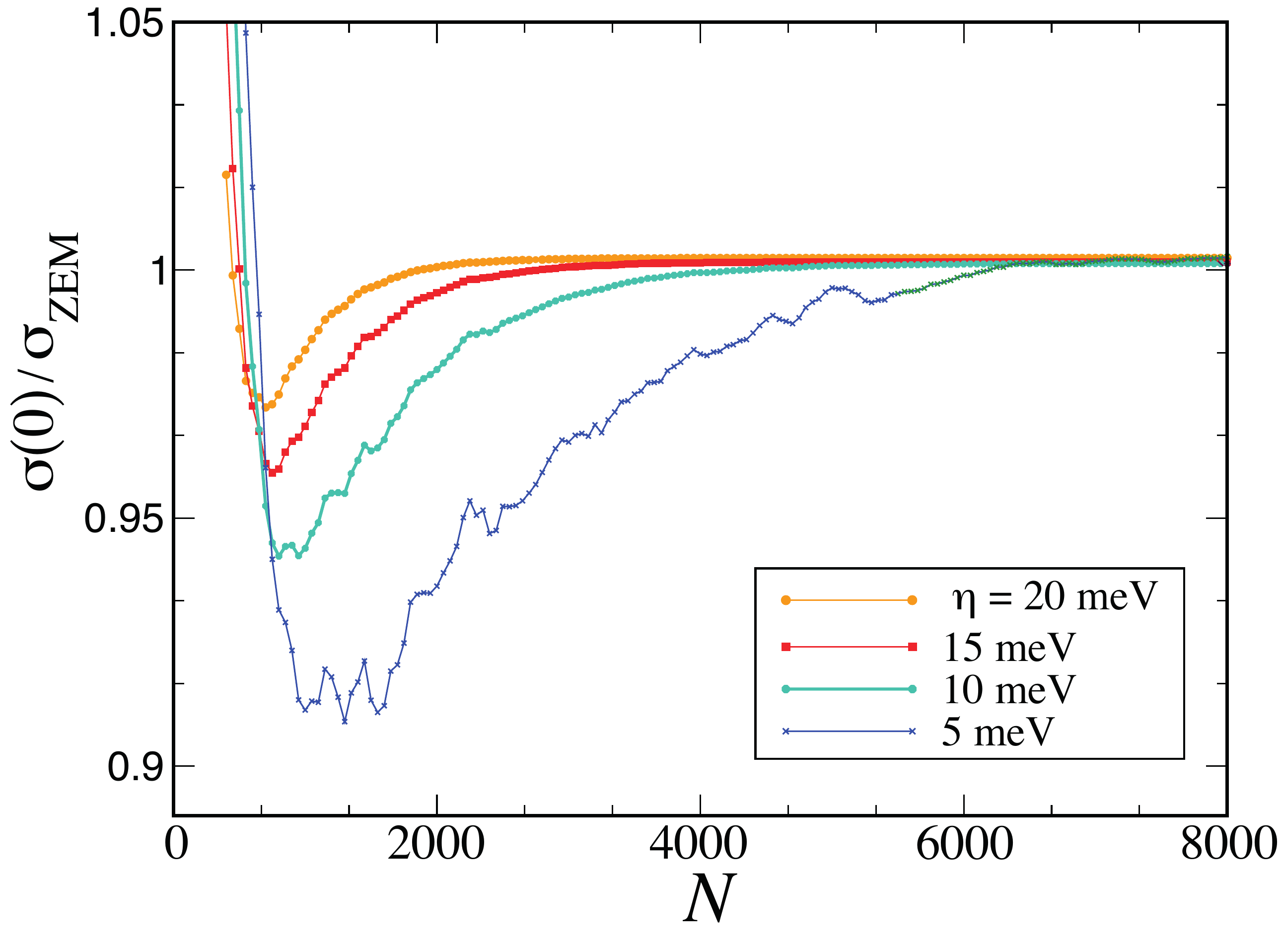} 
\par\end{centering}

\protect\protect\protect\caption{\label{fig:COND_1}Analysis of full-spectral results. \textbf{Top
panel}. Conductivity of graphene with vacancy defects ($n_{i}=0.4\%$)
as function of Fermi energy. The resolution of the calculation is
$\eta=$10~meV {[}$\sigma$ is given in units $\sigma_{\textrm{ZEM}}\equiv4e^{2}/(\pi h)${]}.
The energy grids contain 1000 points. The inset shows the estimated
error based on the standard deviation of 20 independent sets (each
containing an average over 250 random vectors). For comparison, the
error estimated using the electron-hole asymmetry degree is also shown.
For clarity the grid in the inset contains only 49 points with $E>0$.
\textbf{Bottom Panel}. The convergence of the $N$-order approximation
to the ZEMs microscopic conductivity $\sigma_{N}(0,\eta)$ is shown
at selected values of $\eta$. Clearly, several thousands Chebyshev
iterations (corresponding to tens of millions expansion moments $\mathcal{V}_{nm}$)
are required as $\eta$ enters the meV range.}
\end{figure}

\begin{figure}[H]
\begin{centering}
\includegraphics[width=0.85\columnwidth]{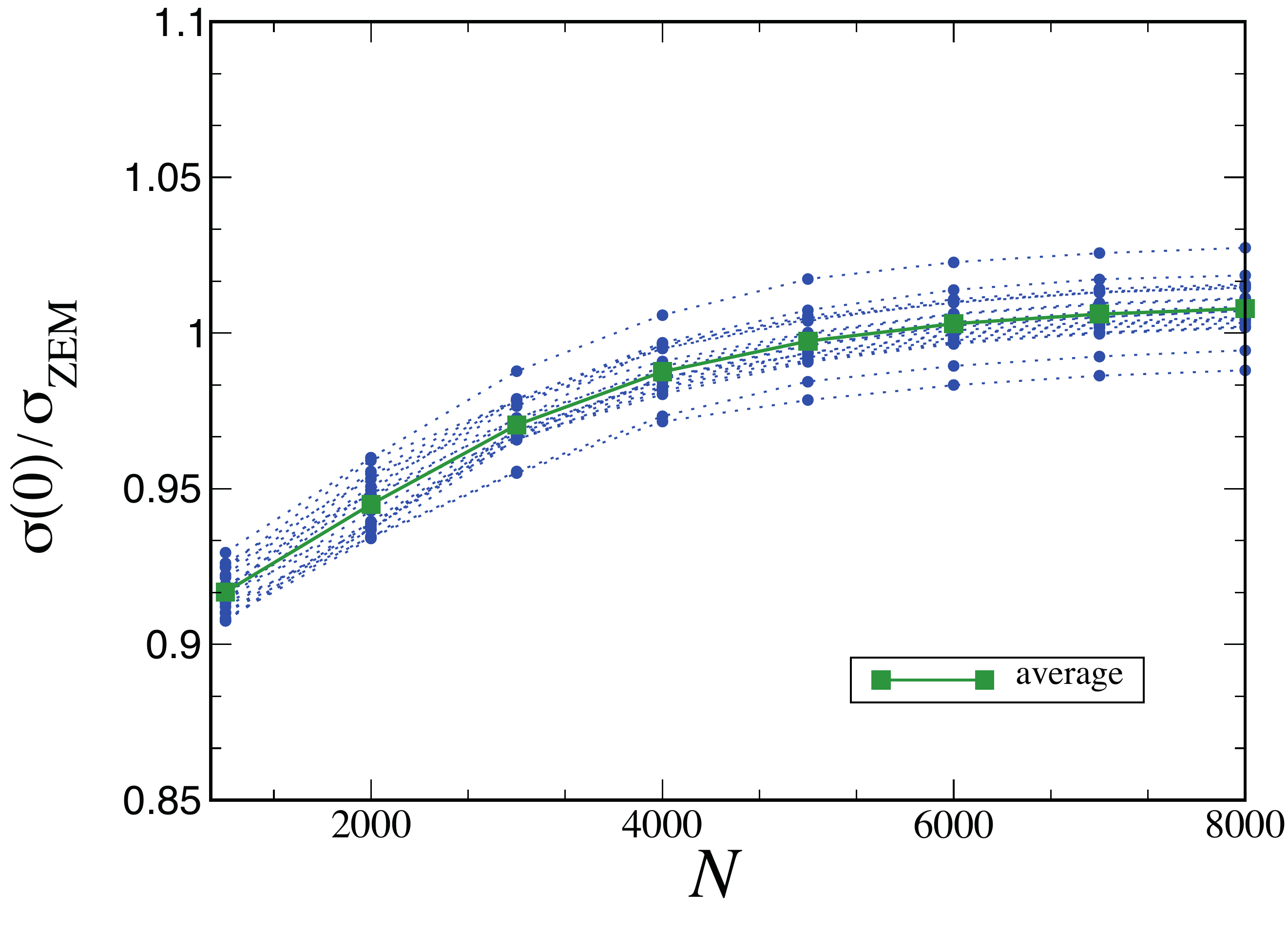} 
\par\end{centering}

\begin{centering}
\includegraphics[width=0.85\columnwidth]{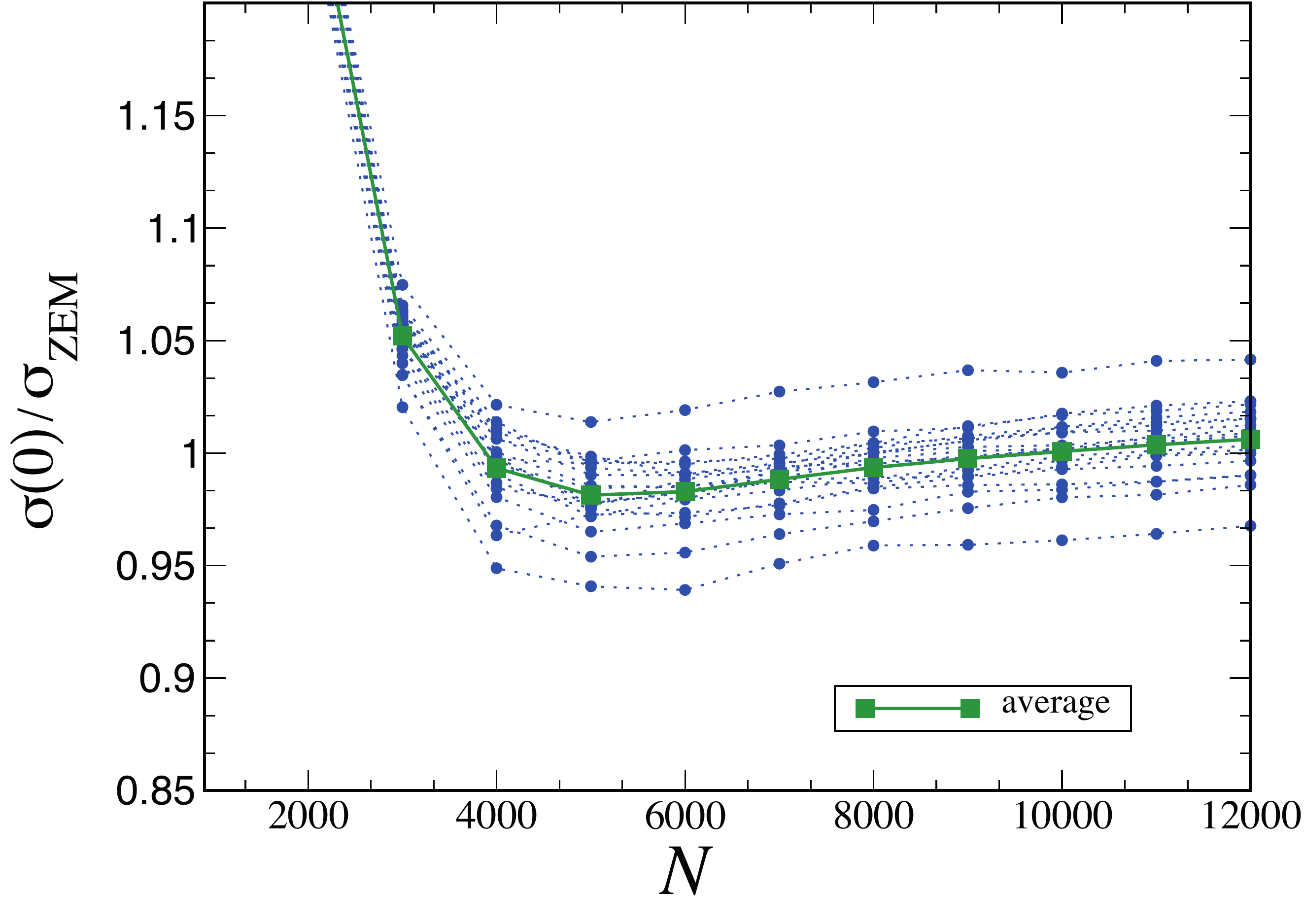} 
\par\end{centering}

\protect\protect\protect\caption{\label{fig:COND_2} Numerical analysis of the large-scale calculations.
\textbf{Top panel}. ``Single-shot'' ZEM conductivity of realistic
size graphene samples with a dilute concentration of vacancies ($n_{i}=0.05\%$)
as function of $N$. The resolution of the calculation is $\eta=$5~meV.
Each data set (blue circles) corresponds to a single system realization.
The average over 20 independent system realizations is shown in squares.
\textbf{Bottom Panel}. The same as in the left panel for $N=12000$
polynomials and a resolution $\eta=$2.5~meV.}
\end{figure}


The results reported in this section are for a honeycomb lattice with
$D=3200\times3200$ sites and a vacancy concentration $n_{i}=0.4\%$.
In order to extract the Kubo conductivity with satisfactory resolution
we computed $N=8000$ Chebyshev iterations {[}corresponding to $N^{2}=6.4\times10^{7}$
moments in the expansion of $\sigma_{N}(E,\eta)$, Eq.~(\ref{eq:Kubo_N}){]}.
The resulting $N\times N$ matrix is subsequently used to evaluate
$\sigma_{N}(E,\eta)$ on a fine grid. $D$ is large enough so that
the thermodynamic limit $\Omega\rightarrow\infty$ can be safely extrapolated.
In Fig.~\ref{fig:COND_1}~(a) we show $\sigma(E)\equiv\lim_{\Omega\rightarrow\infty}\sigma_{N\rightarrow\infty}(E,\eta=10\,\textrm{meV})$
for a fixed disorder realization. Here, $E$ is the Fermi energy in
eV.

Remarkably, the stochastic trace in Eq.~(\ref{eq:STE_COND}) required
thousands random vectors to converge $\sigma_{N}(E,\eta)$ to a good
precision \cite{footnote}. The high degree of electron--hole symmetry
$\sigma(E,\eta)=\sigma(-E,\eta)$ achieved {[}see Fig.~\ref{fig:COND_1}~(a){]}
testifies to the high quality of the results. The error in $\sigma_{N}(E,\lambda)$
is estimated to be in the range 0.1--1\%. This is further confirmed
with a detailed numerical study summarized in the inset to Fig.~\ref{fig:COND_1}
(see caption for details).

The convergence of the $N$-order approximation for ZEMs is shown
in Fig.~\ref{fig:COND_1}~(b). Whereas for poor resolutions $\approx20$~meV
a few thousand Chebyshev iterations are sufficient, probing resolutions
$\approx1$~meV is manifestly more demanding. Moreover, statistical
fluctuations in the STE become important at small $\eta$, which requires
more random vector realizations (see the noise in the curve for $\eta=5$~meV).
Importantly, all the curves studied converge to $\sigma_{\textrm{ZEM}}$
to 1\% accuracy, the main result of the Letter. A dedicated calculation
at $E=0$ will confirm this (see below).


\subsubsection{Single-energy high-resolution results}

In Sec.~\ref{sec:Efficient Recursive Method} we devised a ``single-energy
algorithm'' that bypasses the computation of Chebyshev moments $\mathcal{V}_{nm}$,
allowing us to reach much larger system sizes. We now describe its
application to the problem of the ZEMs in graphene. The calculations
summarized in this section are for a honeycomb lattice with $D=3.6\times10^{9}$
sites. The huge system dimension results in $\sigma(E,\eta)$ data
with satisfactory accuracy even for a single system realization, i.e.,
one random vector $R=1$ and a single (vacancy) disorder realization.
This situation is computationally very convenient as it provides a
quick ``single-shot'' evaluation of the dc conductivity.

In Fig.~\ref{fig:COND_2} we show the variation of $\sigma_{N}(0,\eta)$
with $N$. As mentioned, a single system realization converges $\sigma_{N}(0,\eta)$
to a very reasonable precision (note that the vertical axis is zoomed
around $\sigma=\sigma_{\textrm{ZEM}}$). The error bars increase slowly
with $N$ as the matrices $\mathcal{T}_{n}(\hat{h})$ become less
and less sparse as $n\rightarrow N-1$ for $N\gg1$. For a dilute
vacancy concentration $n_{i}=0.05\%$ and broadening $\eta=5$~meV,
we obtain $\langle\overline{\sigma_{N=8000}(0)}\rangle=1.008$ (in
units of $\sigma_{\textrm{ZEM}}$) with standard deviation $\delta\sigma=0.009$
(corresponding to 0.8\% of the mean value). For the high resolution
calculations ($\eta=2.5$~meV), these values are $\langle\overline{\sigma_{N=12000}(0)}\rangle=1.006$
and 0.016, respectively.

In the main text, a set of ``single-shot'' calculations with $\eta=\{2.5,5,7.5,10,12.5,15,20,40,60\}$~meV
and $n_{i}=0.4\%$ (Fig.~3), and $\eta=5$~meV and $n_{i}=\{0.05,0.1,0.2,0.4,0.6,0.8,1\}\%$
(Fig.~4) were presented. In order obtain a conservative estimate
of the error bars involved we performed 20 independent realizations
of the more disordered system, i.e., $n_{i}=1\%$. We obtained $\langle\overline{\sigma_{N=8000}(0)}\rangle=1.014$
with standard deviation 0.007, which suggests an accuracy of $\approx1$\%.

Probing resolutions resolutions approaching 1~meV becomes increasingly
more challenging as the number of iterations $N$ increases considerably,
and hence the number of random vectors necessary to converge the STE.
We performed a small set of simulations for $\eta=1$~meV ($N=12000$,
averaged over 2 disorder realizations and 10 random vectors) and obtained
an average $0.95\sigma_{\textrm{ZEM}}$ with $5\%$ standard deviation.


\begin{thebibliography}{10}
\bibitem{50YoAL} \emph{50 Years of Anderson Localization}, edited
by E. Abrahams (World Scientific, Singapore, 2010).

\bibitem{Gade=00003D00003D000026Wegner91-93} R. Gade and F. Wegner,
Nucl. Phys. B \textbf{360}, 213 (1991); R. Gade, Nucl. Phys. B \textbf{398},
499 (1993).

\bibitem{Altland=00003D00003D000026Zirnbauer} M. R. Zirnbauer, J.
Math. Phys. \textbf{37}, 4986 (1996); A. Altland and M. R. Zirnbauer,
Phys. Rev. B \textbf{55}, 1142 (1997).

\bibitem{EversMirlin_RMP} F. Evers and A. D. Mirlin, Rev. Mod. Phys.
\textbf{80}, 1355 (2008).

\bibitem{RMPgraphene} A. H. Castro Neto, F. Guinea, N. M. R. Peres,
K. S. Novoselov, and A. K. Geim, Rev. Mod. Phys. \textbf{81}, 109
(2009).

\bibitem{PeresRMP} N. M. R. Peres, Rev. Mod. Phys. \textbf{82}, 2673
(2010).

\bibitem{MuccioloLewenkopf} E. R. Mucciolo and C. H. Lewenkopf, J.
Phys.: Conden. Matter \textbf{22}, 273201 (2010).

\bibitem{Katsnelson06} M. I. Katsnelson, Eur. Phys. J. B \textbf{51},
157 (2006).

\bibitem{Ando02} H. Suzuura and T. Ando, Phys. Rev. Lett.\textbf{
89}, 266603 (2002).

\bibitem{Markos12} L. Schweitzer and P. Marko\v{s}, Phys. Rev. B
\textbf{85}, 195424 (2012).

\bibitem{Balseiro14} G. Usaj, P. S. Cornaglia, and C. A. Balseiro,
Phys. Rev. B \textbf{89}, 085405 (2014).

\bibitem{Evers_DoS_2014} V. Hafner, J. Schindler, N. Weik, T. Mayer,
S. Balakrishnan, R. Narayanan, S. Bera, and F. Evers, Phys. Rev. Lett.
\textbf{113}, 186802 (2014).

\bibitem{Mirlin_DoS_2014} P. M. Ostrovsky, I. V. Protopopov, E. J.
Konig, I. V. Gornyi, A. D. Mirlin, and M. A. Skvortsov, Phys. Rev.
Lett. \textbf{113}, 186803 (2014).

\bibitem{Pereira_07} V. M. Pereira, F. Guinea, J. M. B. Lopes dos
Santos, N. M. R. Peres, and A. H. Castro Neto. Phys. Rev. Lett. \textbf{96},
036801 (2006).

\bibitem{Pereira_08} V. M. Pereira, J. M. B. Lopes dos Santos, and
A. H. Castro Neto, Phys. Rev. B \textbf{77}, 115109 (2008).

\bibitem{Ugeda2010}M. M. Ugeda, I. Brihuega, F. Guinea, and J. M.
Gomez-Rodriguez, Phys. Rev. Lett. \textbf{104}, 096804 (2010).

\bibitem{HHernandoPRL09} D. Huertas-Hernando, F. Guinea, and A. Brataas,
Phys. Rev. Lett. \textbf{103}, 146801 (2009).

\bibitem{Ferreira11} A. Ferreira, J. Viana-Gomes, J. Nilsson, E.
R. Mucciolo, N. M. R. Peres, and A. H. Castro Neto, Phys. Rev. B \textbf{83},
165402 (2011).

\bibitem{Harju14} Z. Fan, A. Uppstu, and A. Harju, Phys. Rev. B \textbf{89},
245422 (2014).

\bibitem{Mayou_2013} G. Trambly de Laissardiere, and D. Mayou, Phys.
Rev. Lett. \textbf{111}, 146601 (2013).

\bibitem{Roche_2013}A. Cresti, F. Ortmann, T. Louvet, D. Van Tuan,
and S. Roche, Phys. Rev. Lett. \textbf{110}, 196601 (2013).

\bibitem{Mirlin_Vacancies_14} S. Gattenlohner, W.-R. Hannes, P. M.
Ostrovsky, I. V. Gornyi, A. D. Mirlin, and M. Titov, Phys. Rev. Lett.
\textbf{112}, 026802 (2014).

\bibitem{Mirlin_Vacancies_10} P. M. Ostrovsky, M. Titov, S. Bera,
I. V. Gornyi, and A. D. Mirlin, Phys. Rev. Lett. \textbf{105}, 266803
(2010).

\bibitem{Mirlin_2006} P. M. Ostrovsky, I. V. Gornyi, and A. D. Mirlin,
Phys. Rev. B \textbf{74}, 235443 (2006).

\bibitem{KPM} A. Weisse, G. Wellein, A. Alvermann, and H. Fehske,
Rev. Mod. Phys. \textbf{78}, 275 (2006).

\bibitem{Boyd} J. P. Boyd, \textit{Chebyshev and Fourier Spectral
Methods}, 2nd edition (Dover Publications, 2001).

\bibitem{Supp_Mat}See Supplemental Material attached below (pages
6-12) for details on the Chebyshev-polynomial Green function (CPGF)
method as well as a thorough description of the accurate large-scale
numerical calculations presented in the Letter, which includes Refs.~\cite{CPGF,Kosloff84,Gradshteyn,Ebisuzaki_2004}.

\bibitem{CPGF}A. Ferreira (to be published).

\bibitem{Kosloff84}H. Tal-Ezer and R. Kosloff, J. Chem. Phys. \textbf{81},
3967 (1984).

\bibitem{Gradshteyn}I. S. Gradshteyn and I. M. Ryzhik. \emph{Table
of Integrals, Series, and Products}, 7th Edition (Elsevier, Academic
Press, 2007).

\bibitem{Ebisuzaki_2004}T. Iitaka and T. Ebisuzaki, Phys. Rev. E
\textbf{69}, 057701 (2004).

\bibitem{Thouless} D. J. Thouless and S. Kirkpatrick, Phys. C: Solid
State Phys. \textbf{14}, 235 (1981).

\bibitem{Imry} Y. Imry, \textit{Introduction to mesoscopic physics},
2nd edition (Oxford University Press, 2002).

\bibitem{Resources} A single realization of a system with $D=3.6\times10^{9}$
atoms with $N$ up to 12000 takes only a few days of HPC time if enough
memory (typically 0.5 TB) is available to recursively construct the
huge vectors $|\varphi_{\pm}(E)\rangle$ on the fly.

\bibitem{MirlinPRB85} E. J. Konig, P. M. Ostrovsky, I. V. Protopopov,
and A. D. Mirlin, Phys. Rev. B \textbf{85}, 195130 (2012).

\end{thebibliography}

\section*{References}
\begin{thebibliography}{10}
\bibitem{CPGF-1}A. Ferreira, unpublished (2015).

\bibitem{rescaling}Without loss of generality we have assumed a bound
spectrum with $|\epsilon_{m}|\le1$. The latter can always be achieved
by rescaling $\hat{H}\rightarrow\hat{h}=(\hat{H}-a_{+}\mathbb{I})/a_{-}$
where $a_{\pm}=\frac{1}{2}\left(\textrm{max}\: E_{m}\pm\textrm{min}\: E_{m}\right)$.

\bibitem{Boyd-1}J. P. Boyd, \textit{Chebyshev and Fourier Spectral
Methods}, second revised edition (Dover Publications, 2001).

\bibitem{Kosloff84-1}H. Tal-Ezer and R. Kosloff, J. Chem. Phys. \textbf{81},
3967 (1984).

\bibitem{Gradshteyn-1}I. S. Gradshteyn and I. M. Ryzhik. \emph{Table
of Integrals, Series, and Products}, 7th Edition (Elsevier, Academic
Press, 2007).

\bibitem{KPM-1}A. Weisse, G. Wellein, A. Alvermann, and H. Fehske,
Rev. Mod. Phys \textbf{78}, 275 (2006).

\bibitem{Unified_MLG_BLG} A. Ferreira, J. Viana-Gomes, J. Nilsson,
E. R. Mucciolo, N. M. R. Peres, and A. H. Castro Neto, Phys. Rev.
B \textbf{83} 165402 (2011).

\bibitem{Imry-1}Y. Imry, \textit{Introduction to mesoscopic physics},
2nd edition (Oxford University Press, 2002).

\bibitem{Ebisuzaki_2004-1}T. Iitaka and T. Ebisuzaki, Phys. Rev.
E \textbf{69}, 057701 (2004).

\bibitem{RAM}For a graphene system with 10 million atoms $D=10^{7}$,
and $N=8\times10^{3}$ recursive steps, one needs $\approx$0.5~TB
of RAM just to store all these vectors since a double precision number
takes 8 bytes.

\bibitem{GadeWegner91-93}R. Gade and F. Wegner, Nucl. Phys. B \textbf{360},
213 (1991); R. Gade, Nucl. Phys. B \textbf{398}, 499 (1993).

\bibitem{Evers_2014}V. Häfner, J. Schindler, N. Weik, T. Mayer, S.
Balakrishnan, R. Narayanan, S. Bera, and F. Evers, Phys. Rev. Lett.
\textbf{113}, 186802 (2014).

\bibitem{Mirlin_2014}P. M. Ostrovsky, I. V. Protopopov, E. J. König,
I. V. Gornyi, A. D. Mirlin, and M. A. Skvortsov, Phys. Rev. Lett.
\textbf{113}, 186803 (2014).

\bibitem{footnote}Usually it is assumed that the fluctuations in
the stochastic trace evaluation are of the order of $1/\sqrt{RD}$,
such that for large $D$ a few random vectors are required \cite{KPM,Ebisuzaki_2004}.
This result follows from the sparseness of the operators involved.
However, in general the quality of the stochastic trace is very sensitive
to the number of Chebyshev iterations. This happens because for large
$n$ (typically $n\approx\sqrt{D}$) the Chebyshev operator $T_{n}(\hat{h})$
is no longer sparse. To the best of our knowledge, this fact has remained
unnoticed in previous works.\end{thebibliography}
\end{document}